\documentclass[aps,prc,preprintnumbers,amsmath,amssymb,superscriptaddress,floatfix,groupedaddress,showpacs]{revtex4-1}
\usepackage{epsfig}% Include figure files
\usepackage{dcolumn}% Align table columns on decimal point
\usepackage{bm}% bold math   % for math
\usepackage{amssymb}
\begin{document}
\title{Hadron Mass Spectrum and the Shear Viscosity to Entropy Density Ratio of Hot Hadronic Matter}

\author{Jacquelyn Noronha-Hostler}
\author{Jorge Noronha}
\affiliation{Instituto de F\'{i}sica, Universidade de S\~{a}o Paulo, C.P.
66318, 05315-970 S\~{a}o Paulo, SP, Brazil}

\author{Carsten Greiner}
\affiliation{Institut f\"ur Theoretische Physik, Goethe Universit\"at, Frankfurt, Germany}

%\date{\today}
\begin{abstract}
Lattice calculations of the QCD trace anomaly at temperatures $T<160$ MeV have been shown to match hadron resonance gas model calculations, which include an exponentially rising hadron mass spectrum. 
In this paper we perform a more detailed comparison of the model calculations to lattice data that confirms the need for an exponentially increasing density of hadronic states. Also, we find that the lattice data is compatible with a hadron density of states
 that goes as $\rho(m) \sim m^{-a}\exp(m/T_H) $ at large $m$ with $a> 5/2$ (where $T_H \sim 167$ MeV). With this specific subleading contribution to the density of states, heavy resonances are most likely undergo 2-body decay (instead of multi-particle decay), 
which facilitates their inclusion into hadron transport codes. Moreover, estimates for the shear viscosity and the shear relaxation time coefficient of the hadron resonance model computed within the excluded volume approximation suggest that these 
transport coefficients are sensitive to the parameters that define the hadron mass spectrum. 
\end{abstract}
\maketitle

\section{Introduction}

Particle flow anisotropies at low transverse momentum produced in ultrarelativistic heavy ion collisions can be reasonably described \cite{bjorn,song,romatschke} using relativistic fluid dynamics with a very small shear viscosity
to entropy ratio $\eta/s \sim 1/(4\pi)$. This is as small as the uncertainty principle-based estimate derived by Danielewicz and Gyulassy nearly 30 years ago \cite{Danielewicz:1984ww} and also the more 
recent calculations \cite{Kovtun:2004de} performed in strongly-coupled gauge theories dual to higher dimensional theories of gravity \cite{maldacena}. Beyond leading log perturbative QCD calculations that are applicable
at temperatures $T>m_{pion}$ give values for the ratio that are an order of magnitude larger than the bound \cite{Arnold:2000dr} (for calculations based on parton transport see \cite{greinerxu}). Moreover, calculations 
performed using hadronic models at $T\sim m_{pion}$ also resulted in values for the ratio above the viscosity bound \cite{Gorenstein:2007mw,Itakura:2007mx,bass,pal}. 

Due to the observation made in \cite{kapusta1} that a small value of 
$\eta/s$ in QCD should occur in the transition region $T\sim 150-200$ MeV due to the rapid increase in the entropy density observed in lattice simulations \cite{Cheng:2009zi,Borsanyi:2010cj}, the effects of an exponentially 
increasing density of hadronic states on several properties of hot hadronic matter were investigated using the hadron resonance gas model in Refs.\ 
\cite{NoronhaHostler:2007jf,NoronhaHostler:2008ju,NoronhaHostler:2009tz,NoronhaHostler:2009hp,NoronhaHostler:2009cf}. It was shown in those studies that the addition of 
new hadronic states that follow an exponentially increasing hadron mass spectrum as proposed by Hagedorn \cite{Hagedorn:1965st} 
\begin{equation}\label{hagedornspectrum}
\lim_{m\to \infty}\rho(m) \sim \frac{e^{m/T_H}}{m^{5/2}}
\end{equation}
to the hadron resonance gas model led to a much better agreement to the lattice data computed around the pseudo-critical QCD critical temperature $T_c\sim 196$ MeV inferred from Ref.\ \cite{Cheng:2007jq}. Moreover, an estimate
of $\eta/s$ at $T\sim 190$ MeV computed using this model indicated that excited hadronic matter at those temperatures could become a nearly perfect fluid \cite{Schafer:2009dj} where $\eta/s$ approached $1/(4\pi)$. In this
case, the transition from viscous hydrodynamics to typical hadronic transport would be much smoother than expected. However, with the advent of the lattice calculations published in \cite{Borsanyi:2010cj} 
and the smaller critical region $T\sim 155$ MeV obtained in that study (which has been independently confirmed in \cite{Bazavov:2011nk}), a revision of the effects of heavy resonances on the hadron 
resonance gas model became necessary. Ref.\ \cite{Majumder:2010ik} showed that a hadron resonance gas model containing only the known hadrons and resonances could only describe the lattice data up to $T\sim 140$ MeV while the 
inclusion of states with mass $m>2$ GeV (which follow an exponential spectrum) could improve the match to the lattice data of \cite{Borsanyi:2010cj} and provide a good description of lattice QCD thermodynamics up to $T= 155$ MeV.

In this paper we present a more detailed comparison between the hadron resonance gas model calculations and the lattice data that not only provides strong evidence for the need of an exponentially increasing density of 
hadronic states with mass $m>2$ GeV but also indicates that the density of states goes as $\rho(m) \sim m^{-a}\exp(m/T_H)/$ at large $m$ with $a>5/2$ (where $T_H \sim 167$ MeV). As in \cite{Majumder:2010ik}, we estimate that
the maximum temperature at which the hadron resonance gas model is applicable is $\sim 155$ MeV. A rough estimate of the shear viscosity computed within the excluded volume approximation for the hadron resonance 
model suggests that the shear viscosity to entropy ratio of hot hadronic matter is sensitive to the parameters that describe the hadron mass spectrum. We briefly comment
also on the value of the shear relaxation time coefficient of hot hadronic matter.

\section{Hadron Mass Spectrum and the Hadron Resonance Gas Model}

In the statistical bootstrap model of hadrons \cite{Hagedorn:1965st,Frautschi:1971ij}, a hadron is considered to be a  volume $V$ (with typical length $\sim 1$ fm) composed of two or more freely roaming constituents 
and the hadron density of states is required to be consistent with the spectrum of constituents, which are themselves hadrons. This is the so-called ``bootstrap condition'' pionered in this context by Hagedorn in 1965 
\cite{Hagedorn:1965st}. Frautschi \cite{Frautschi:1971ij} reformulated the bootstrap condition and wrote down an equation for the total density of states $\rho(m)$ in the hadronic volume (for Boltzmann statistics) where different
 states of mass $m_i$ and energy $E_i=\sqrt{p_i^2+m_i^2}$ inside the box possess single particle density $\rho_{in}(m_i)$. The bootstrap condition  
\begin{equation}
 \lim_{m\to \infty}\rho(m)\Longrightarrow \rho_{in}(m)
\end{equation}
is exactly satisfied when $\lim_{m\to \infty}\rho(m)\sim c\, e^{bm}/m^a$ with $a>5/2$ \cite{Frautschi:1971ij}. This exponentially rising mass spectrum is typical of a system consisting of string-like constituents 
such as a gas of free strings \cite{veneziano,Huang:1970iq,Cohen:2006qd} or large $N_c$ glueballs \cite{Thorn:1980iv}. The spectrum of experimentally measured hadrons \cite{Eidelman:2004wy} was found to be 
compatible with an exponential increase of the number of states up to $\sim 1.7$ GeV \cite{Broniowski:2000bj,Broniowski:2004yh} although the subleading power $a$ cannot be reliably determined from such an analysis. 

Once the hadron mass spectrum is given, the thermodynamical quantities of the hadron resonance gas model (at zero chemical potential) in the total volume $V$ and temperature $T$ are fully determined by the partition function 
(assuming Boltzmann statistics)
\begin{equation}
 Z(T,V) = \sum_{N=0}^\infty \frac{1}{N!}\prod_{i=1}^N \int dm_i\, \rho(m_i) \int \frac{d^3 p_i}{(2\pi)^3}e^{-E_i/T} V^N
\label{partitionfunction}
\end{equation}
which can be used to determine the usual thermodynamic functions. In fact, one finds 
\begin{equation}
 p(T) = \frac{T^2}{2\pi^2} \int_{0}^\infty dm\,\rho(m) m^2 K_2\left(\frac{m}{T}\right) \\
\label{pressure1}
\end{equation}
for the pressure,
\begin{equation}
 \varepsilon(T) = \frac{1}{2\pi^2} \int_{0}^\infty dm\,\rho(m) m^4 \left[3 \left(\frac{T}{m}\right)^2 K_2\left(\frac{m}{T}\right)+\left(\frac{T}{m}\right) K_1\left(\frac{m}{T}\right)\right]  \\
\label{energydensity1}
\end{equation}
for the energy density,
\begin{equation}
 s(T) = \frac{dp(T)}{dT}=\frac{1}{2\pi^2} \int_{0}^\infty dm\,\rho(m) m^3 K_3\left(\frac{m}{T}\right) \\
\label{entropydensity1}
\end{equation}
for the entropy density, and
\begin{equation}
 \varepsilon(T)-3p(T) = \frac{T}{2\pi^2} \int_{0}^\infty
 dm\,\rho(m) m^3 K_1\left(\frac{m}{T}\right) \\
\label{traceanomaly1}
\end{equation}
for the trace anomaly. The speed of sound can be found from the relation $c_s^2=dp/d\varepsilon$. In this paper we will discuss 4 different forms for the density of states
\begin{eqnarray}
\rho_1(m)&=&A_1\,e^{m/T_{H_1}}\label{rhomueller}\\
\rho_2(m)&=&\frac{A_2}{\left[m^2 +m_{02}^2\right]^{5/4}}\,e^{m/T_{H_2}}\label{rhohagedorn}\\
\rho_3(m)&=&\frac{A_3}{\left[m^2 +m_{03}^2\right]^{3/2}}\,e^{m/T_{H_3}}\label{rhofrautschi} \\
\rho_4(m)&=&\frac{A_4}{T_{H_4}} \left(\frac{m}{T_{H_4}}\right)^\alpha\label{rhoshuryak} 
\end{eqnarray}
where the parameters are shown in Table \ref{tab:par}.  
\begin{table}
\begin{center}
 \begin{tabular}{|c|c|c|c|c|c|c|}
 \hline
  & $T_H$ (GeV) & $A$  & $m_0$ (GeV) & $\alpha$ \\
 \hline
 $\rho_1$ & 0.252 & 2.84 (1/GeV)   &     &   \\
 $\rho_2$ & 0.180 & 0.63 (GeV$^{3/2}$) & 0.5  & \\
 $\rho_3$ & 0.175 & 0.37 (GeV$^{2}$) & 0.5  & \\
 $\rho_4$ & 0.158 & 0.51  &   &  2 \\
 \hline
 \end{tabular}
 \end{center}
 \caption{Parameters for the mass spectra shown in Eqs.\ (\ref{rhomueller})-(\ref{rhoshuryak}).}
 \label{tab:par}
 \end{table}    

\begin{figure}
\centering
\epsfig{file=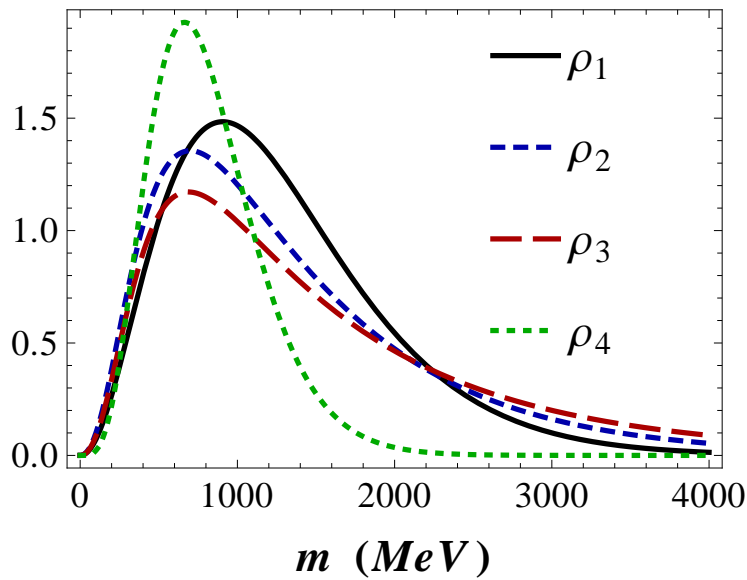,width=0.6\linewidth,clip=}
\caption{Comparison between the integrand (in units of 1/GeV) 
in Eq.\ (\ref{traceanomaly1}) computed using the different $\rho$'s at $T=150$ MeV as a function of $m$. The black line was computed using $\rho_1$, the blue dashed line with $\rho_2$, the red long dashed line with $\rho_3$ while
the short dashed green line was obtained using $\rho_4$.} \label{fig1}
\end{figure}

The parameters chosen for $\rho_1$ and $\rho_2$ are the same used in \cite{Majumder:2010ik} while the parameters for the other $\rho$'s were obtained from a 
fit to the lattice data of Ref.\ \cite{Borsanyi:2010cj}. The density $\rho_3$ satisfies the asymptotic bootstrap condition exactly while $\rho_1$ (introduced in \cite{Majumder:2010ik}) and Hagedorn's $\rho_2$ satisfy
 the bootstrap condition within a power of $m$ \cite{Frautschi:1971ij}. The power law increase given by $\rho_4$ (introduced by Shuryak in the early 70's \cite{Shuryak:1972zq}) does not satisfies the bootstrap condition
 (it also does not lead to any singularities in the thermodynamics) but it provides a nice alternative to describe the rise of the hadronic mass spectrum. As discussed in \cite{Majumder:2010ik}, the 
trace anomaly at temperatures around $160$ MeV becomes sensitive to the heavy states in the spectrum with mass $m>2$ GeV. A comparison between the different $\rho$'s used here can be found in Fig.\ \ref{fig1} where 
the integrand (in units of 1/GeV) in Eq.\ (\ref{traceanomaly1}) is plotted as a function of $m$ at $T=150$ MeV. One can see that the integrand computed using the first three $\rho$'s (black solid, blue dashed, red long dashed, respectively) are very 
similar but they can be clearly distinguished from the result obtained using the power law in $\rho_4$ (small dashed green curve).        

Even though the upper limit of the mass integrals in Eqs.\ (\ref{pressure1}-\ref{traceanomaly1}) is taken to infinity, the divergences implied by an exponentially rising spectrum do not appear in the calculations performed here
because the limiting temperatures $T_{H_i}$'s ($i=1,2,3$) are above the largest temperature considered in this paper $\sim 160$ MeV. One may wonder if the approximations made in (\ref{pressure1}-\ref{traceanomaly1}) (i.e., 
classical statistics and continuous mass spectrum) are at all justified. After all, we know that the measured hadronic spectrum is of course discrete and that baryons and mesons obey different statistics. However, as pointed out
in \cite{Majumder:2010ik}, the simplified formulas in Eqs.\ (\ref{pressure1}-\ref{traceanomaly1}) provide an excellent description of the thermodynamic properties of the hadron resonance gas computed using the measured hadrons in 
the particle data book with the correct statistics in the temperature range $T \sim 100-140$ MeV if one imposes an upper cutoff for the mass integrals. In fact, for $\rho_1$ and $\rho_2$ the mass cutoff is $1.7$ GeV and $1.9$ 
GeV, respectively \cite{Majumder:2010ik}. We have verified that $p/T^4$ computed using $\rho_3$ with a mass cutoff of $1.9$ GeV approaches the result obtained using $\rho_1$ with a mass cutoff of $1.9$ GeV. When
$T < 100$ MeV, the discreteness of the hadron spectrum becomes relevant and the continuous approximation discussed here gives a poor description of the thermodynamic quantities of a hadron resonance gas. Therefore, in order to have a hadronic equation of 
state that is both valid at low temperatures ($T<100$ MeV) and higher temperatures a hybrid model containing the measured hadron states plus a continuous Hagedorn spectrum above a certain mass cutoff is
 more appropriate \cite{NoronhaHostler:2007jf,NoronhaHostler:2008ju,NoronhaHostler:2009tz,NoronhaHostler:2009hp,NoronhaHostler:2009cf}.

\begin{figure}
\centering
\epsfig{file=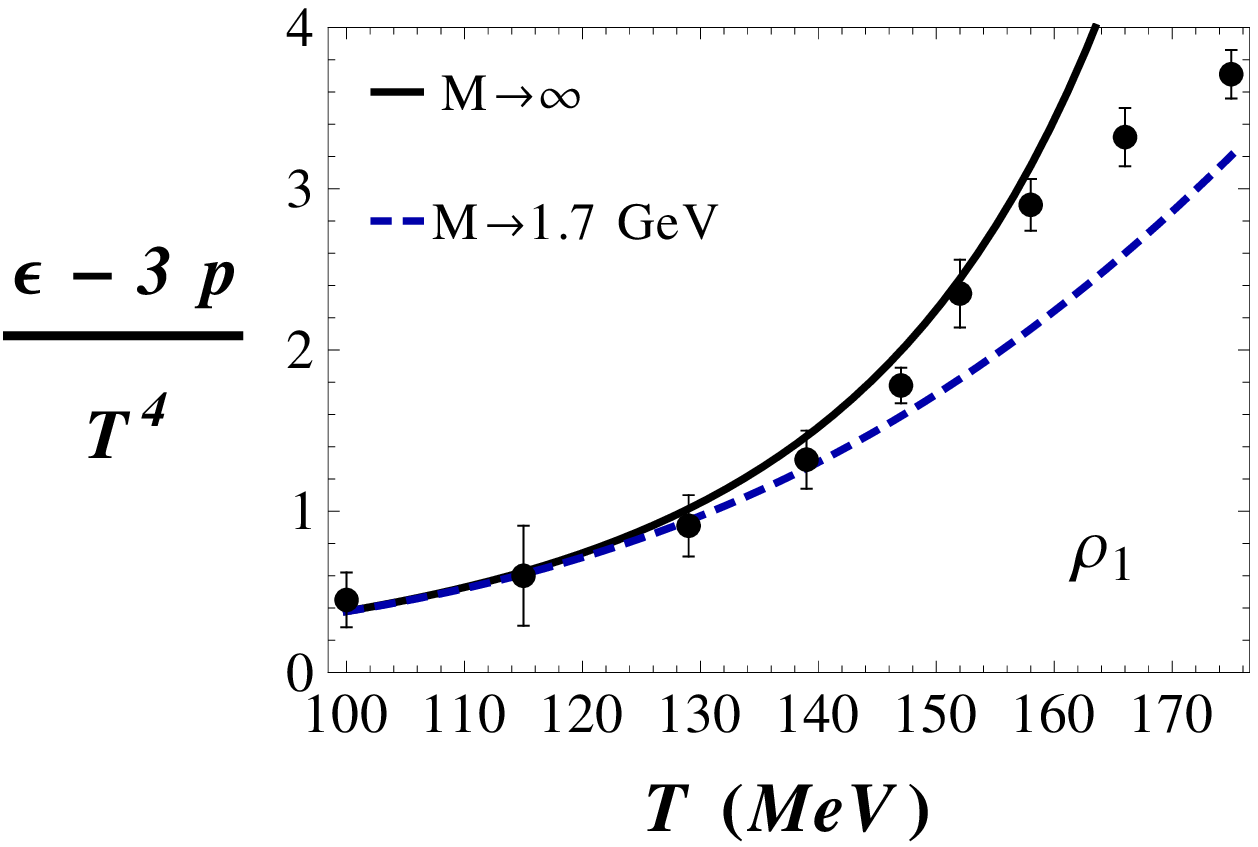,width=0.6\linewidth,clip=}\\
\epsfig{file=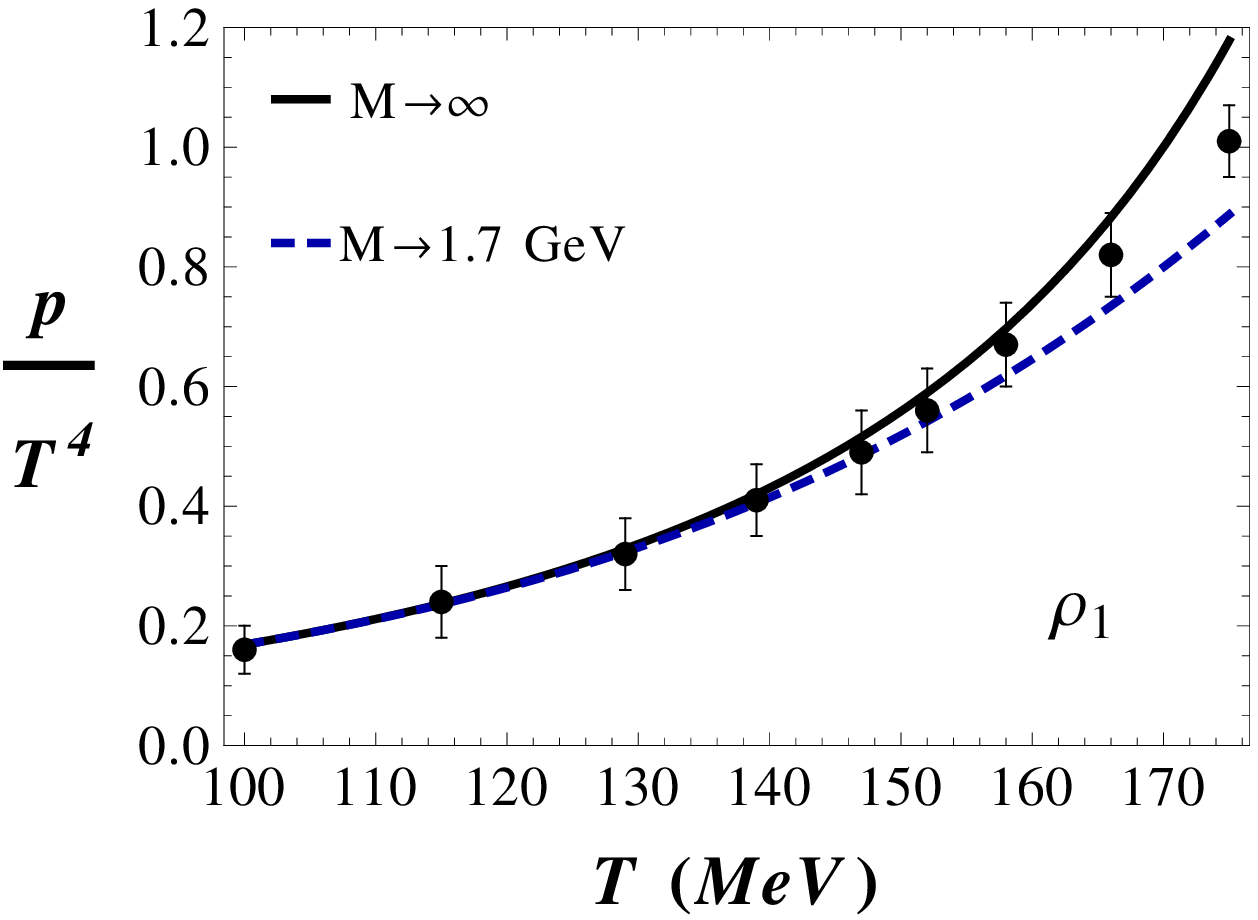,width=0.6\linewidth,clip=}\\
\epsfig{file=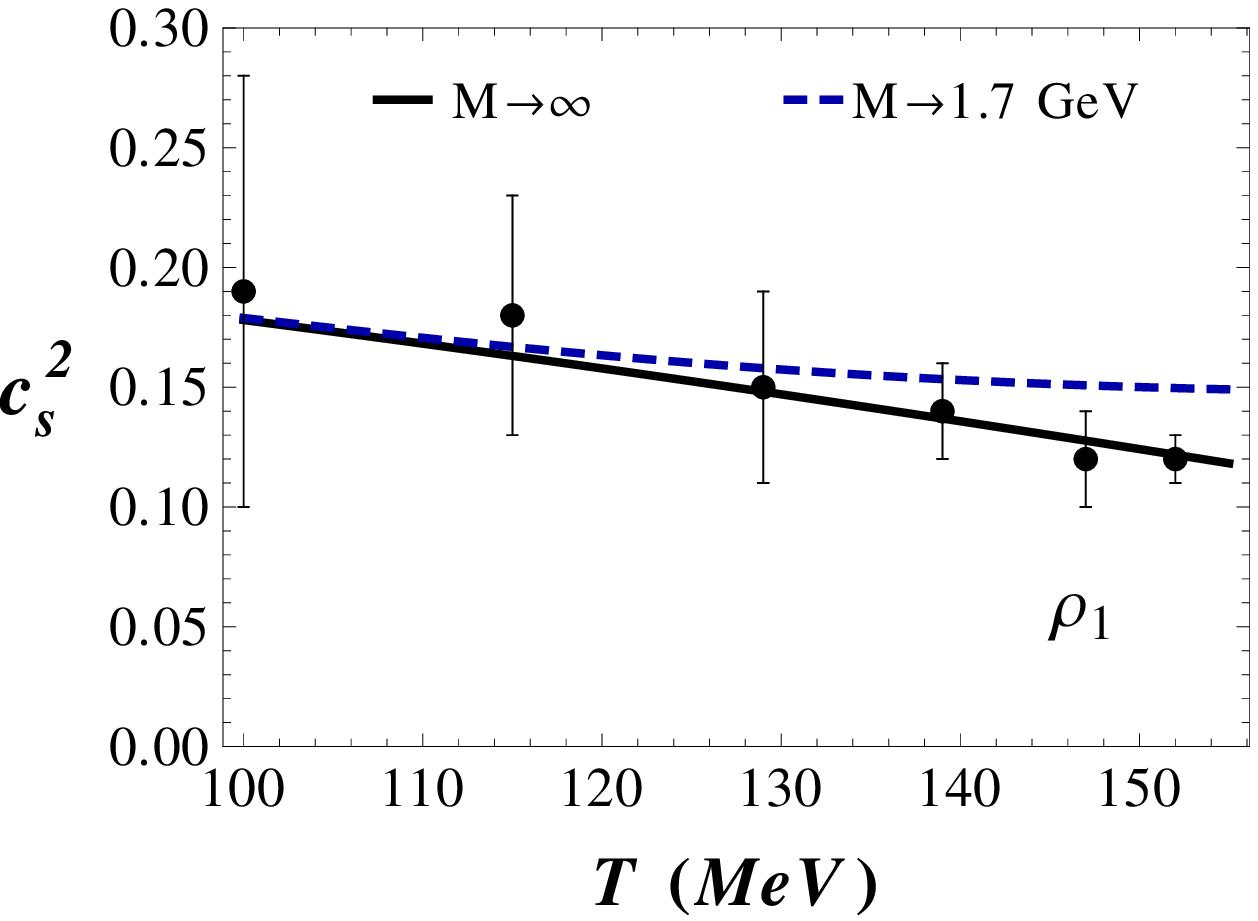,width=0.6\linewidth,clip=}
\caption{Trace anomaly, pressure, and speed of sound squared for the hadron resonance model with density of states $\rho_1$. The black solid curves denote the result obtained by taking the mass integrals in 
Eqs.\ (\ref{pressure1}-\ref{traceanomaly1}) to infinity while the dashed blue curves were computed imposing an upper mass cutoff of 1.7 GeV. The data points correspond to the $N_t=10$ lattice data published in
Ref.\ \cite{Borsanyi:2010cj} (obtained from table 5 in that paper).} \label{thermoplotrho1}
\end{figure}

\begin{figure}
\centering
\epsfig{file=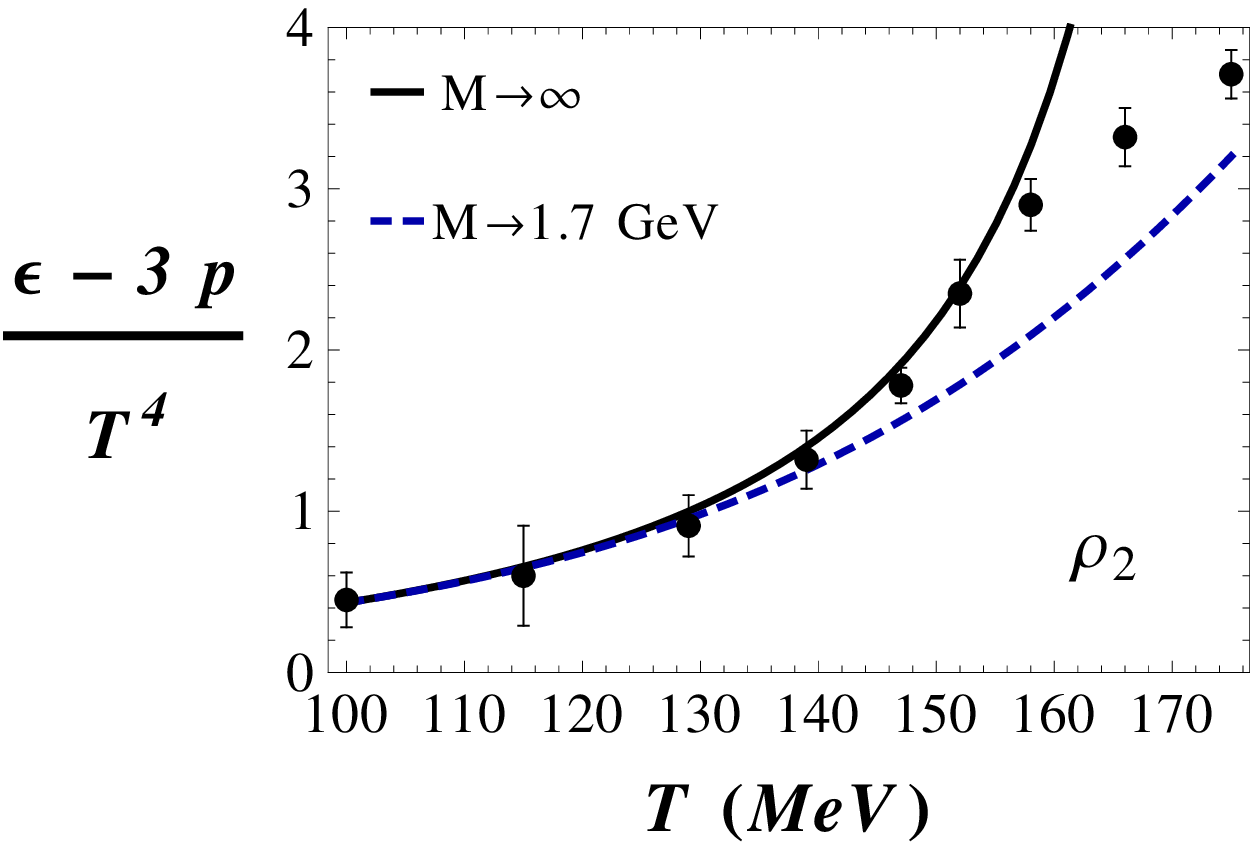,width=0.6\linewidth,clip=}\\
\epsfig{file=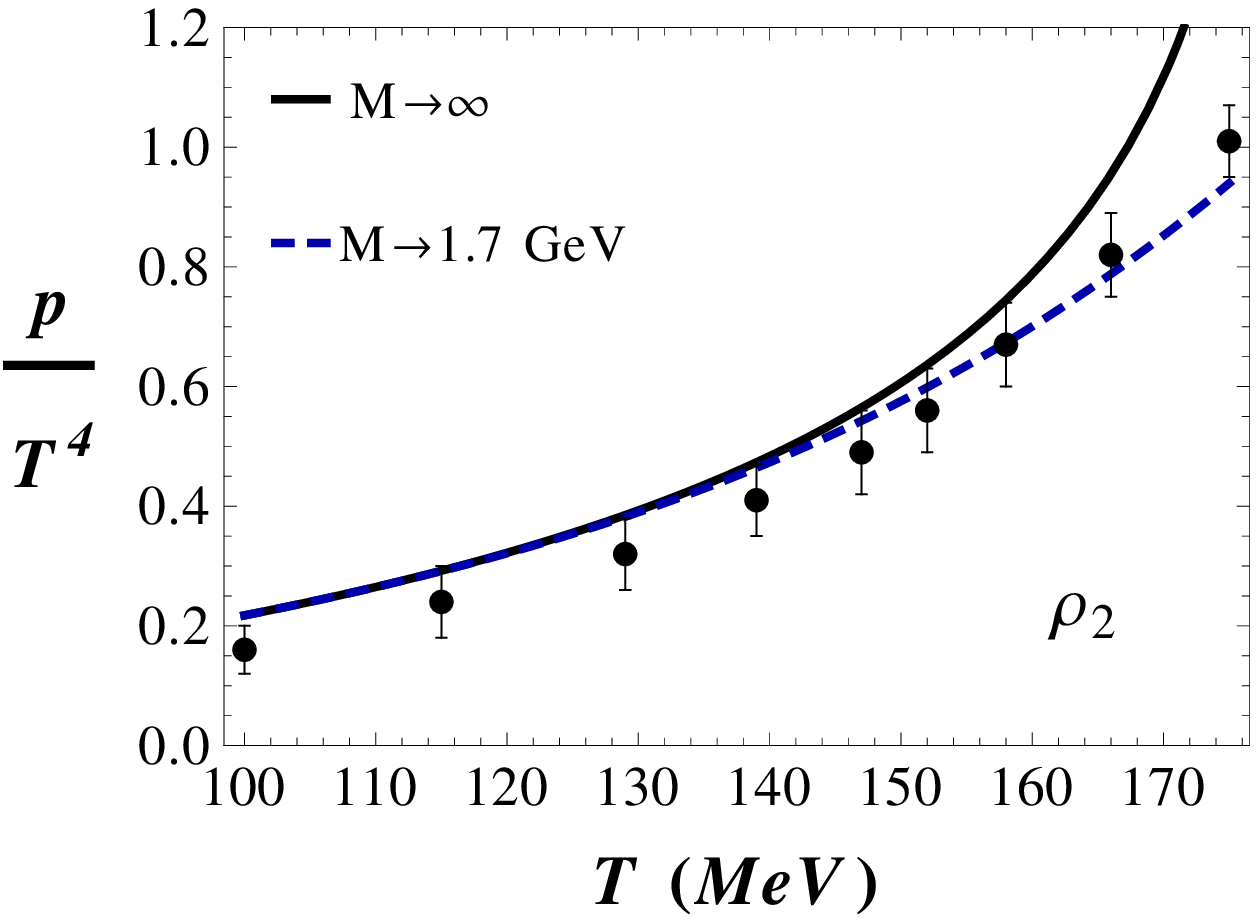,width=0.6\linewidth,clip=}\\
\epsfig{file=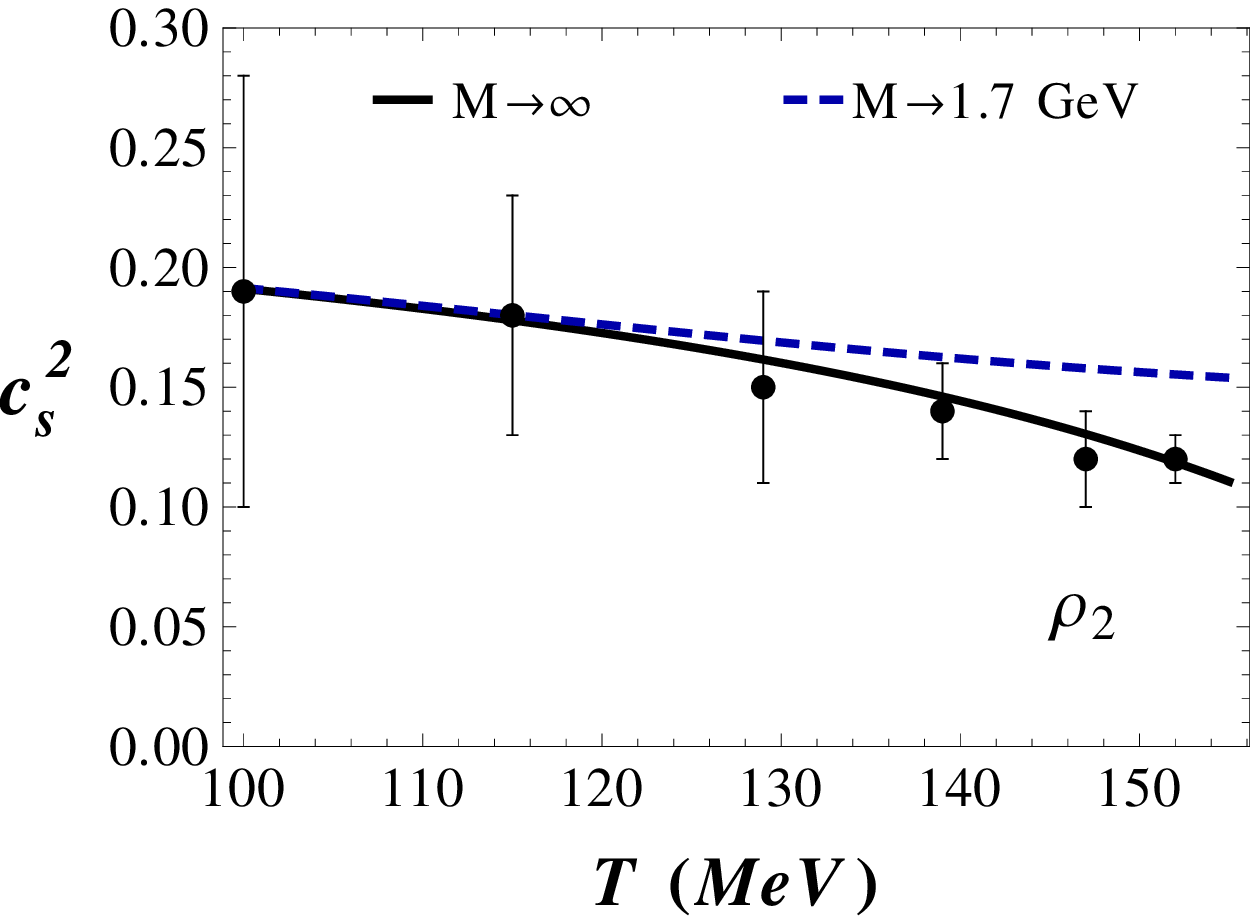,width=0.6\linewidth,clip=}
\caption{Trace anomaly, pressure, and speed of sound squared for the hadron resonance model with density of states $\rho_2$. The black solid curves denote the result obtained by taking the mass integrals in 
Eqs.\ (\ref{pressure1}-\ref{traceanomaly1}) to infinity while the dashed blue curves were computed imposing an upper mass cutoff of 1.9 GeV. The data points correspond to the $N_t=10$ lattice data published in
Ref.\ \cite{Borsanyi:2010cj} (obtained from table 5 in that paper).} \label{thermoplotrho2}
\end{figure}

\begin{figure}
\centering
\epsfig{file=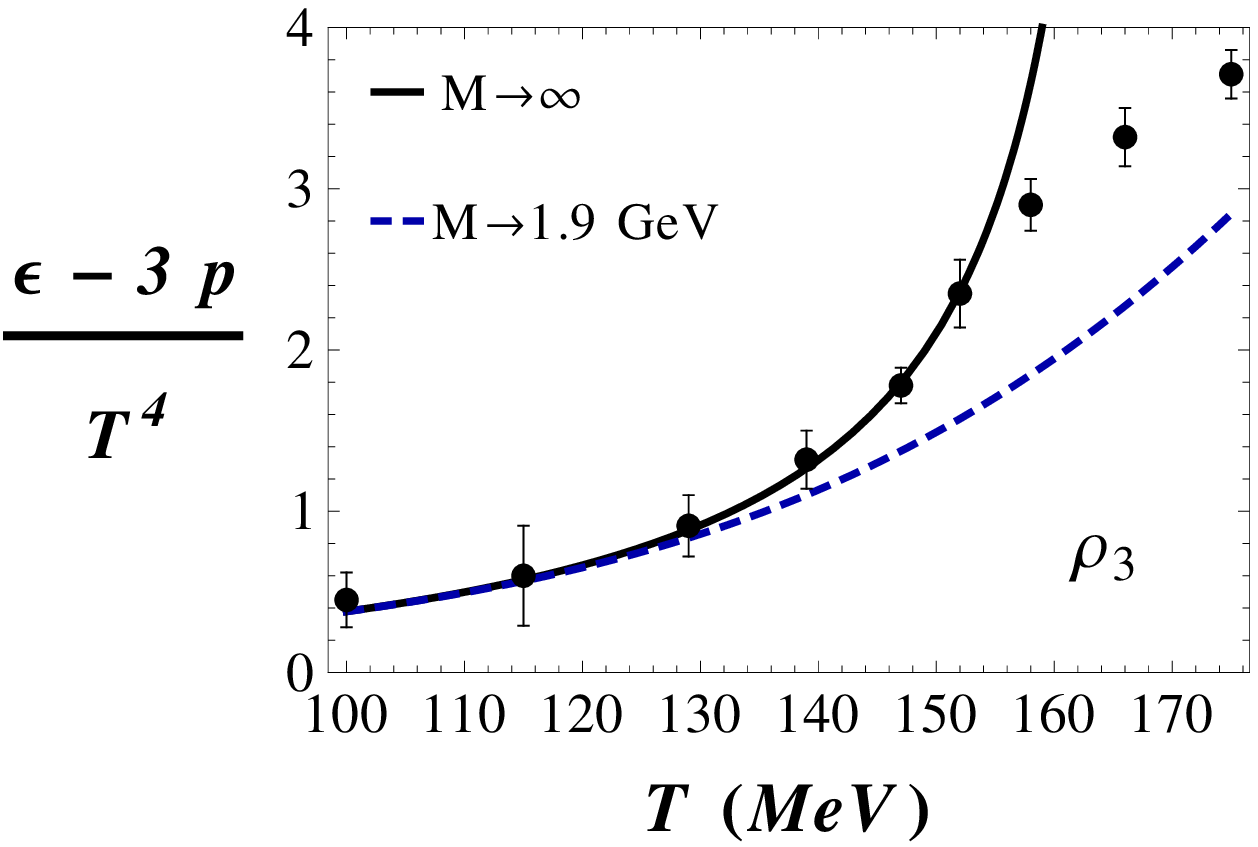,width=0.6\linewidth,clip=}\\
\epsfig{file=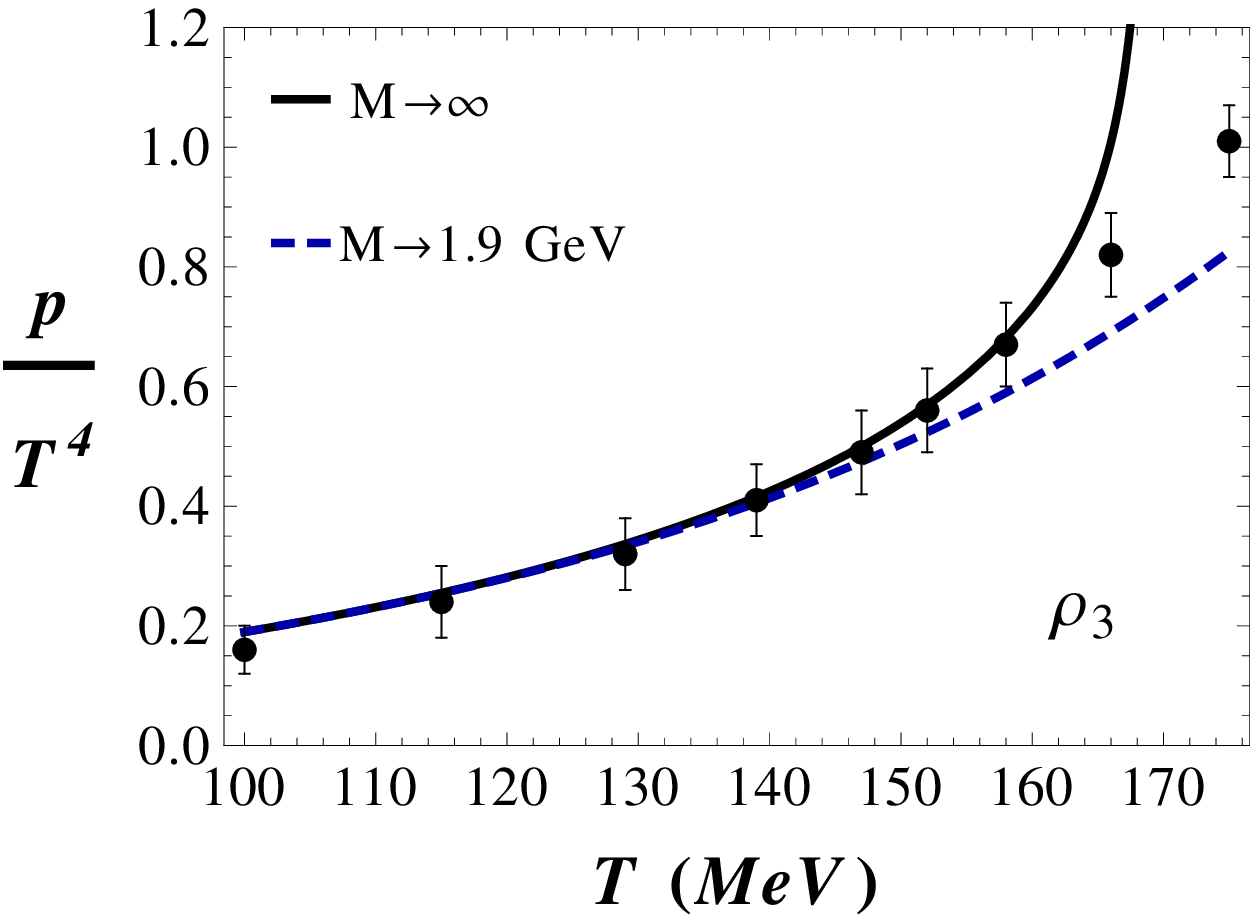,width=0.6\linewidth,clip=}\\
\epsfig{file=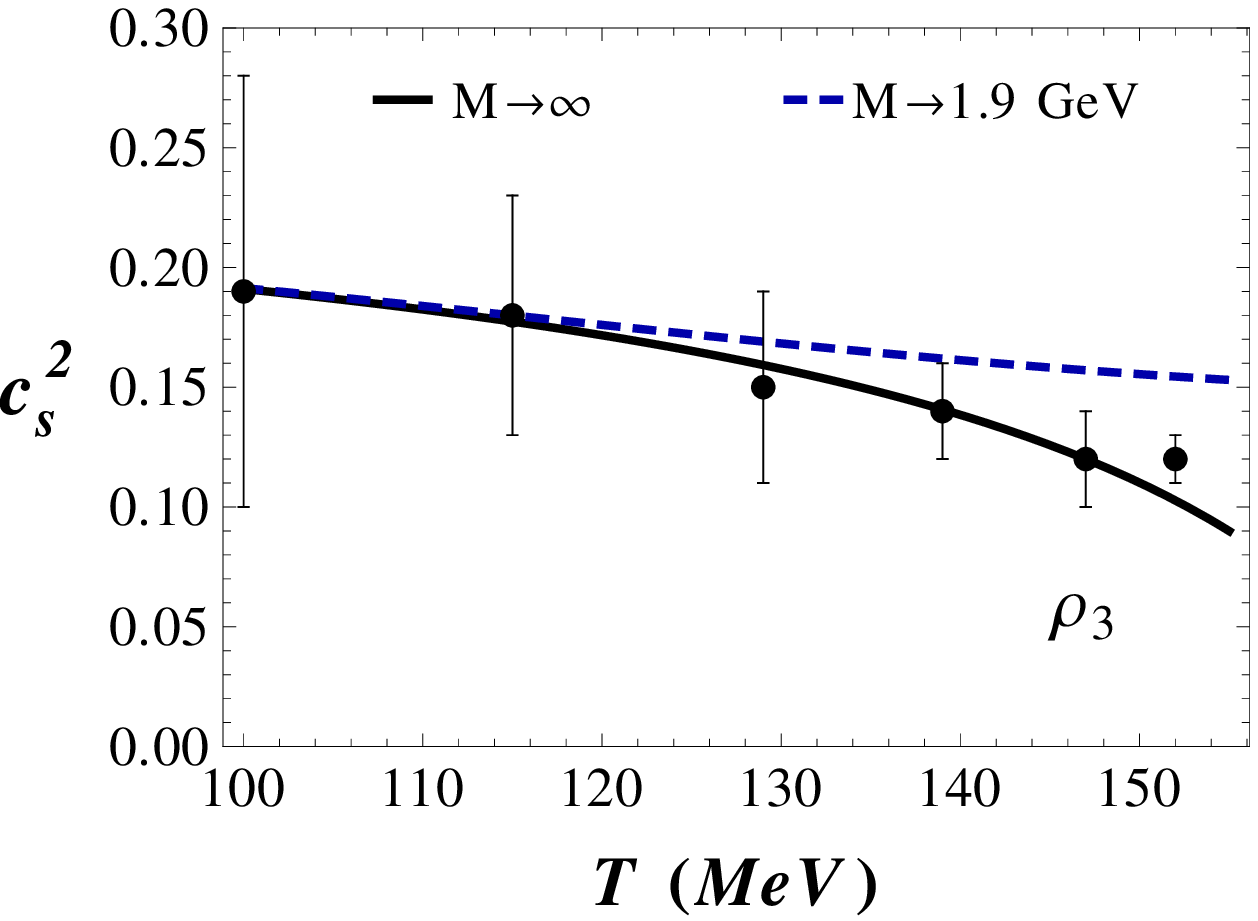,width=0.6\linewidth,clip=}
\caption{Trace anomaly, pressure, and speed of sound squared for the hadron resonance model with density of states $\rho_3$. The black solid curves denote the result obtained by taking the mass integrals in 
Eqs.\ (\ref{pressure1}-\ref{traceanomaly1}) to infinity while the dashed blue curves were computed imposing an upper mass cutoff of 1.9 GeV. The data points correspond to the $N_t=10$ lattice data published in
Ref.\ \cite{Borsanyi:2010cj} (obtained from table 5 in that paper).} \label{thermoplotrho3}
\end{figure}

\begin{figure}
\centering
\epsfig{file=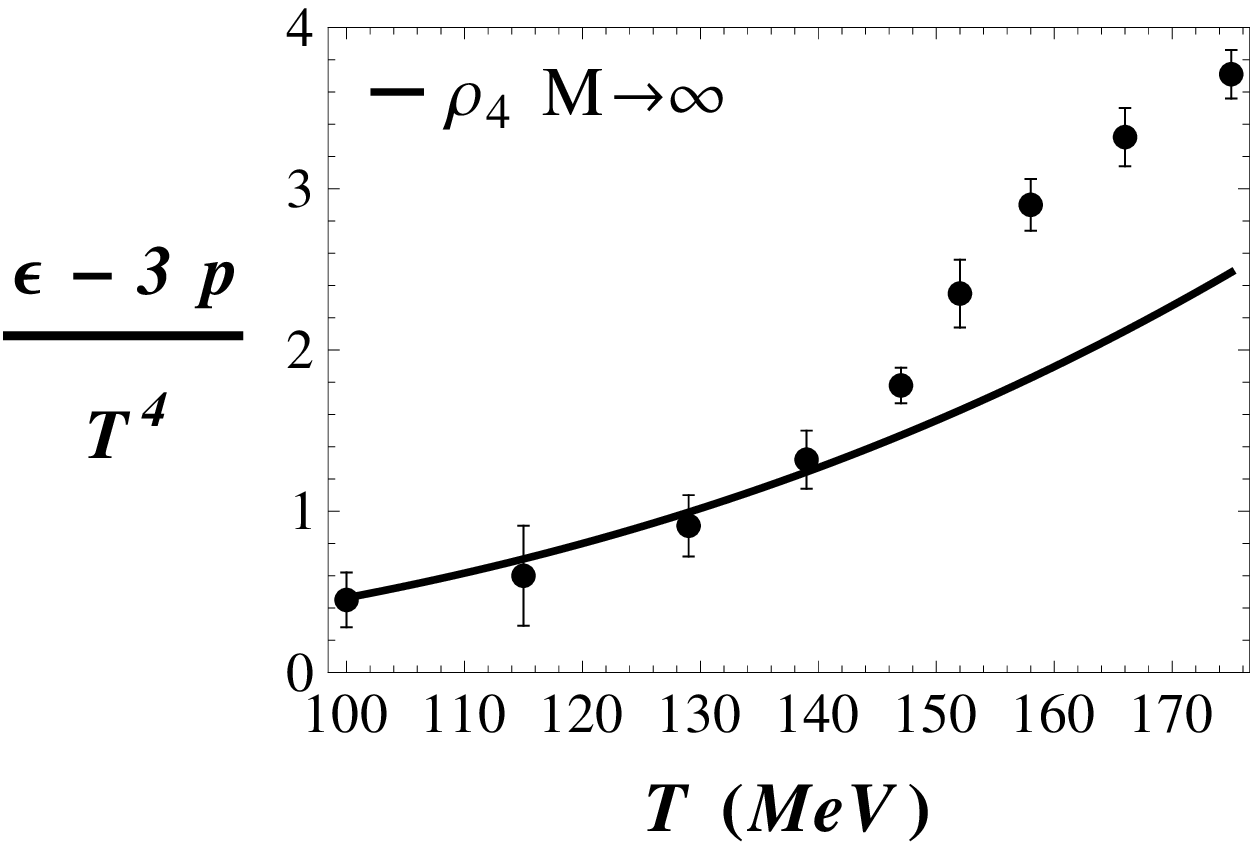,width=0.6\linewidth,clip=}\\
\epsfig{file=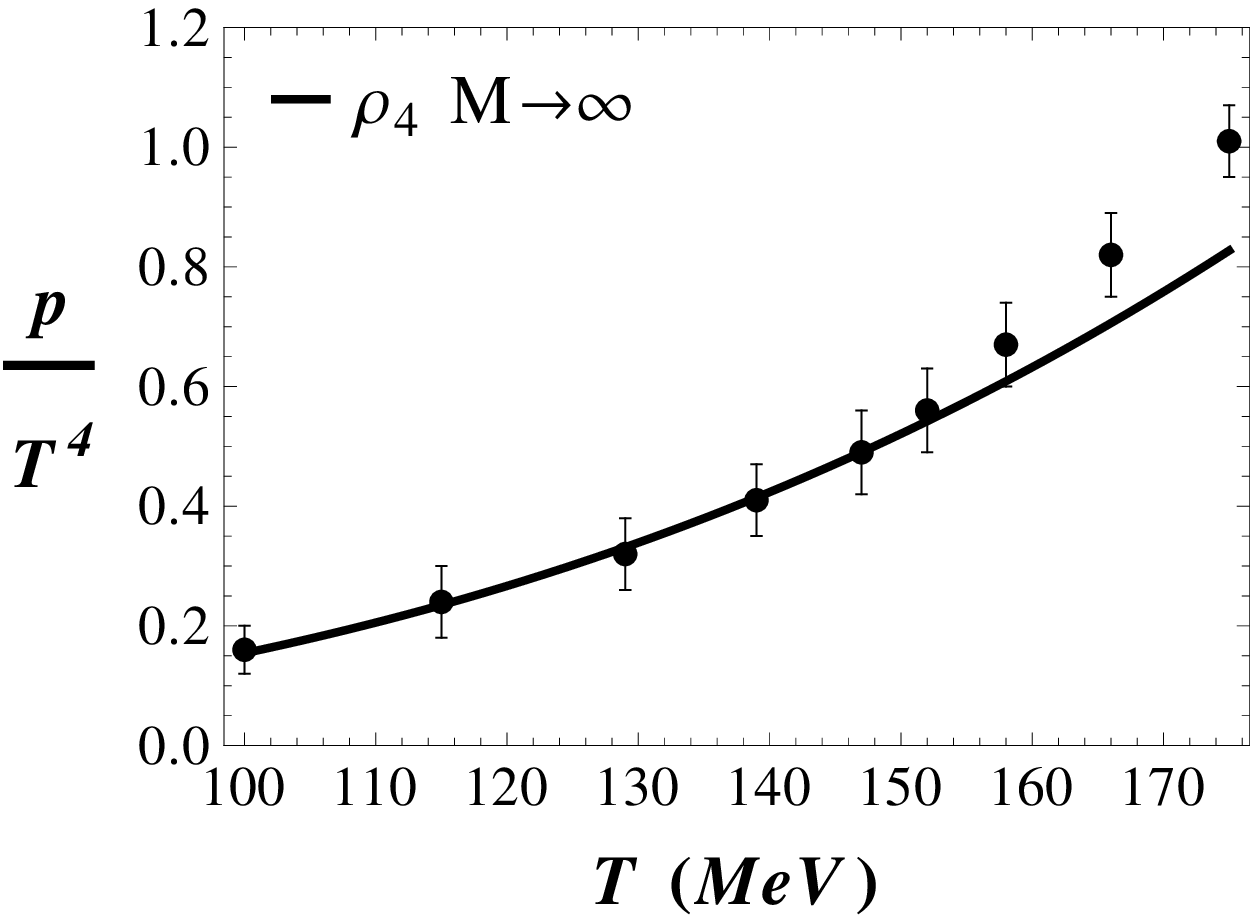,width=0.6\linewidth,clip=}\\
\epsfig{file=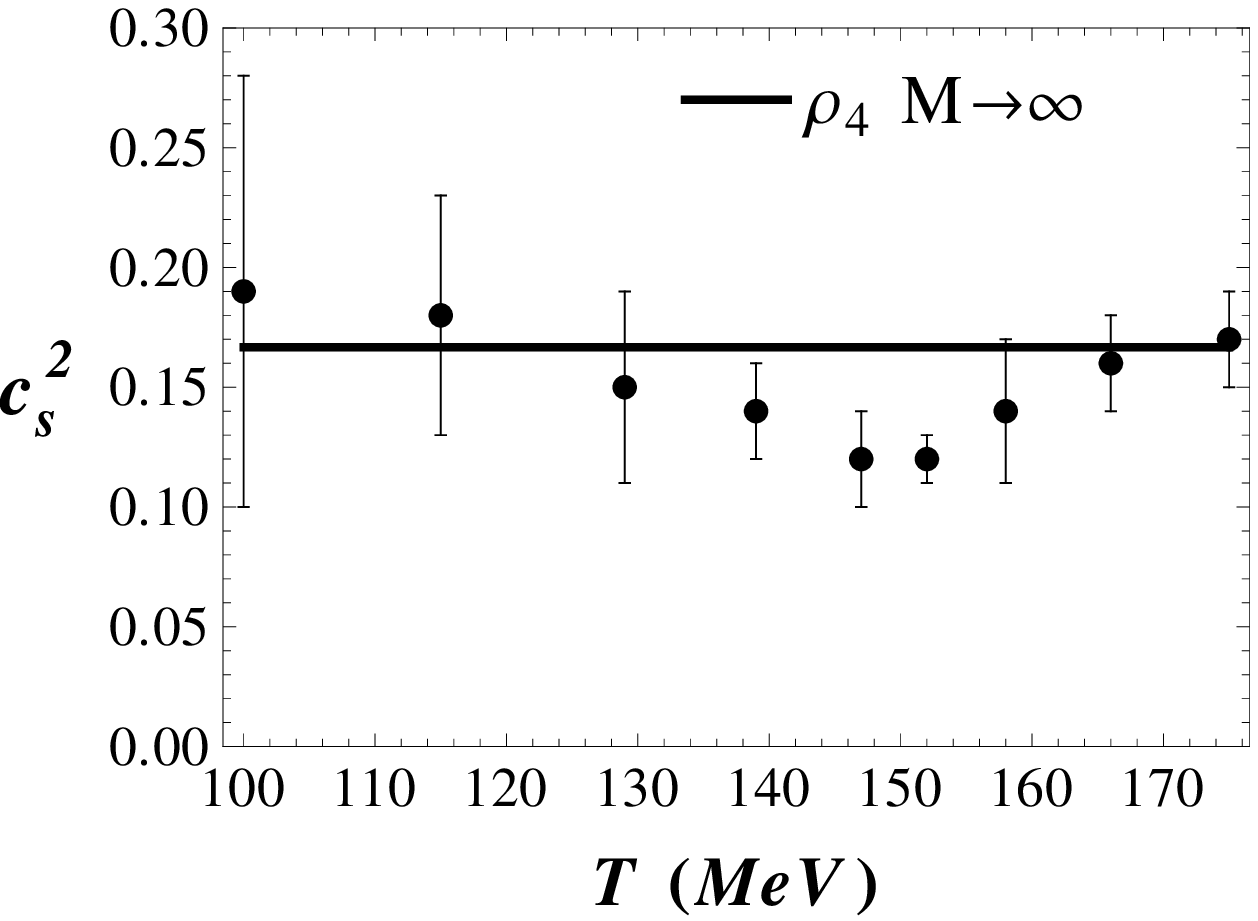,width=0.6\linewidth,clip=}
\caption{Trace anomaly, pressure, and speed of sound squared for the hadron resonance model with density of states $\rho_4$. The black solid curves denote the result obtained by taking the mass integrals in 
Eqs.\ (\ref{pressure1}-\ref{traceanomaly1}) to infinity. The data points correspond to the $N_t=10$ lattice data published in
Ref.\ \cite{Borsanyi:2010cj} (obtained from table 5 in that paper).} \label{thermoplotrho4}
\end{figure}

A comparison between $p(T)/T^4$, $(\varepsilon-3p)/T^4$, and $c_s^2(T)$ of the model defined by Eqs.\ (\ref{pressure1}-\ref{traceanomaly1}) (for the four different hadron density of states) and the $N_t=10$ lattice data of 
Ref.\ \cite{Borsanyi:2010cj} can be found in Figs.\ \ref{thermoplotrho1}-\ref{thermoplotrho4}. The black solid curves denote the result obtained by taking the mass integrals in 
Eqs.\ (\ref{pressure1}-\ref{traceanomaly1}) to infinity while the dashed blue curves were computed imposing an upper mass cutoff that varied for each $\rho$: for $\rho_1$ the cutoff is $1.7$ GeV while for $\rho_2$ and $\rho_3$ the cutoff
is $1.9$ GeV. These cutoffs were determined by requiring that the trace anomaly computed in this continuous model matches the result (up to $T\sim 140$ MeV) obtained in a model where all the hadron resonances of the particle data book are included, as
defined in \cite{Majumder:2010ik}.

The power law increase given by $\rho_4$ considerably simplifies the integrals in (\ref{pressure1}-\ref{traceanomaly1}) and all of them can be done analytically. For instance, one finds in this case that 
$c_s^2 = 1/(\alpha+4)$. However, note in Fig.\ \ref{thermoplotrho4} that the power law spectrum lacks the exponential growth necessary to describe the lattice data for temperatures above 140 MeV. 
This provides evidence that the thermodynamic quantities of QCD computed on the lattice can be understood in terms of a simple hadron resonance gas model with an {\it exponentially rising} density of 
states at temperatures $T\sim 100-155$ MeV. This comparison to lattice data cannot pin down the exact subleading power of $m$ in the hadron density of states. However, it is important to note that the specific value of this power has some interesting 
consequences, as we shall elaborate below.  
 
The subleading contribution $\sim m^{-a}$ to the density of states at large $m$, according to Frautschi's seminal paper \cite{Frautschi:1971ij}, determines the decay properties of a heavy resonance. For instance, when 
$a>5/2$ (which is the case of $\rho_3$), a heavy resonance decays 
(in the first generation of its decay chain) into a heavy secondary particle that carries almost all the available mass plus one (with 69\% probability) or two light hadrons (with 24\% probability). 
This should be contrasted with Hagedorn's original mass spectrum in Eq.\ (\ref{rhohagedorn}) for which the statistically favored
process involves a heavy resonance of mass $m$ decaying into a number $n \sim \ln \,m$ of secondary particles, each of similar mass \cite{Frautschi:1971ij}.   

It is well known that hadrons interact with each other in a variety of different channels, some of them give rise to repulsive interactions while others represent attractive interactions. In Ref.\ \cite{Welke:1990za} 
the pressure of an interacting gas of pions was calculated within the virial expansion (using experimentally determined phase shifts) and it was shown that the thermodynamic quantities of this interacting system nearly coincides 
with those of a free gas of pions and $\rho$ mesons. In this case, there is an approximate cancellation between the attractive and repulsive S-wave channels which effectively enhances the P-wave contribution from
the $\rho$ resonance \cite{Venugopalan:1992hy}. As more hadronic species are included, it is not at all guaranteed that the standard assumption behind hadron resonance models, i.e., that the interacting hadronic 
system can be described by a free gas of the original hadrons and their resonances, is applicable. In general, the inclusion of resonances represents the contribution from the
attractive channels while repulsive interactions can be modeled using simple excluded volume corrections to the thermodynamics \cite{rafelski,kapustaolive,Rischke:1991ke}. The suggestion, obtained 
from a comparison to lattice data, that the complicated interactions among hadrons that enter in the calculations of QCD thermodynamics at temperatures of the order of the pion mass can be effectively modeled 
by a simple, non-interacting gas of hadrons and resonances in accordance with the bootstrap model \cite{Hagedorn:1965st} is therefore quite unexpected and remarkable.

Given the known uncertainties in lattice calculations at low temperatures, the conclusion made above regarding the applicability of the hadron resonance gas should be taken with great care. If results obtained with finer lattices
confirm this picture, this would provide very strong evidence for the validity of Hagedorn's bootstrap hypothesis. We note in passing that recent lattice calculations of the thermodynamical properties 
of $SU(3)$ pure glue have shown evidence for the presence of an exponentially rising glueball mass spectrum below the deconfinement critical temperature \cite{Meyer:2009tq,Borsanyi:2012ve}.

\section{Excluded Volume Corrections to the Hadron Resonance Gas Model}

As mentioned in the previous section, in general one should expect that there are repulsive interactions among hadrons and that a simple way to take that into account in the hadron resonance model is via the excluded volume
corrections \cite{rafelski,kapustaolive,Rischke:1991ke}. In this case, the partition function in (\ref{partitionfunction}) becomes
\begin{equation}
 Z(T,V) = \sum_{N=0}^\infty \frac{1}{N!}\prod_{i=1}^N \int dm_i\, \rho(m_i) \int \frac{d^3 p_i}{(2\pi)^3}e^{-E_i/T} \left(V-\sum_{j=1}^{N}V_j\right)^N
\label{partitionfunctionvdw}
\end{equation}
where $V_j$ denotes the excluded volume by the $j$th hadron. We shall assume for simplicity that the volume excluded by each hadron is a constant that is basically the same for all hadrons, i.e., $V_j=v$. This parameter can be
written in terms of an effective hard-core volume, $v=4 \cdot 4\pi r^3/3$, where $r$ is the effective core radius. The excluded volume pressure is determined by the equation 
\begin{equation}
 \frac{p_v(T)}{T}= n(T) \exp(-vp_v(T)/T)
\label{vdwpressure}
\end{equation}
where $n(T)=p(T)/T$ is the total particle density computed without volume corrections \cite{Rischke:1991ke}. The equation above can be solved analytically in terms of the Lambert $W$ function \cite{comment} and it reads
\begin{equation}
 \frac{p_v(T)}{T}= W(v\, n(T))\,.
\label{vdwpressuresolution}
\end{equation}
The other thermodynamic quantities, $\varepsilon_v(T)$, $s_v(T)$, and $n_v(T)$ can be obtained through the pressure using the standard thermodynamic identities. In the limit where $v\to 0$ one recovers the formulas in 
(\ref{pressure1}-\ref{traceanomaly1}).

\begin{figure}
\centering
\epsfig{file=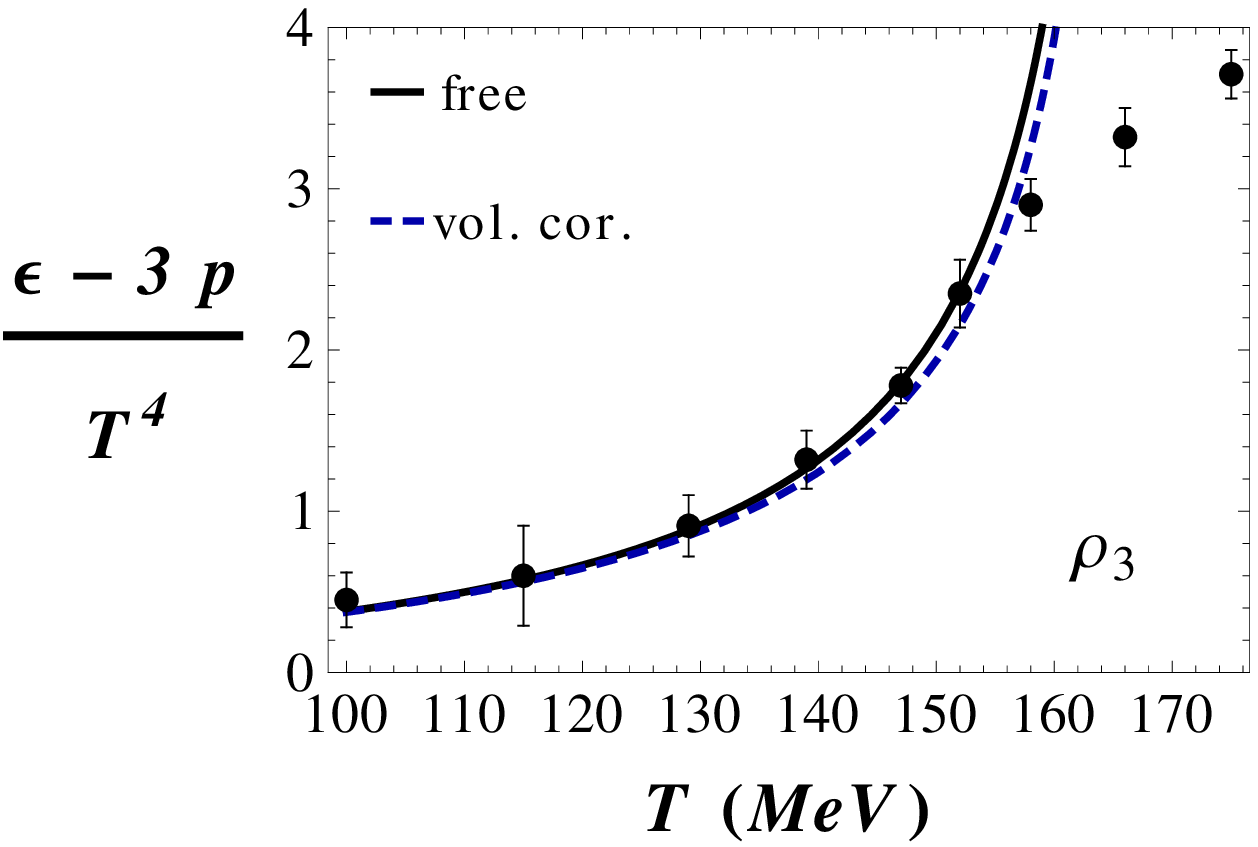,width=0.6\linewidth,clip=}\\
\epsfig{file=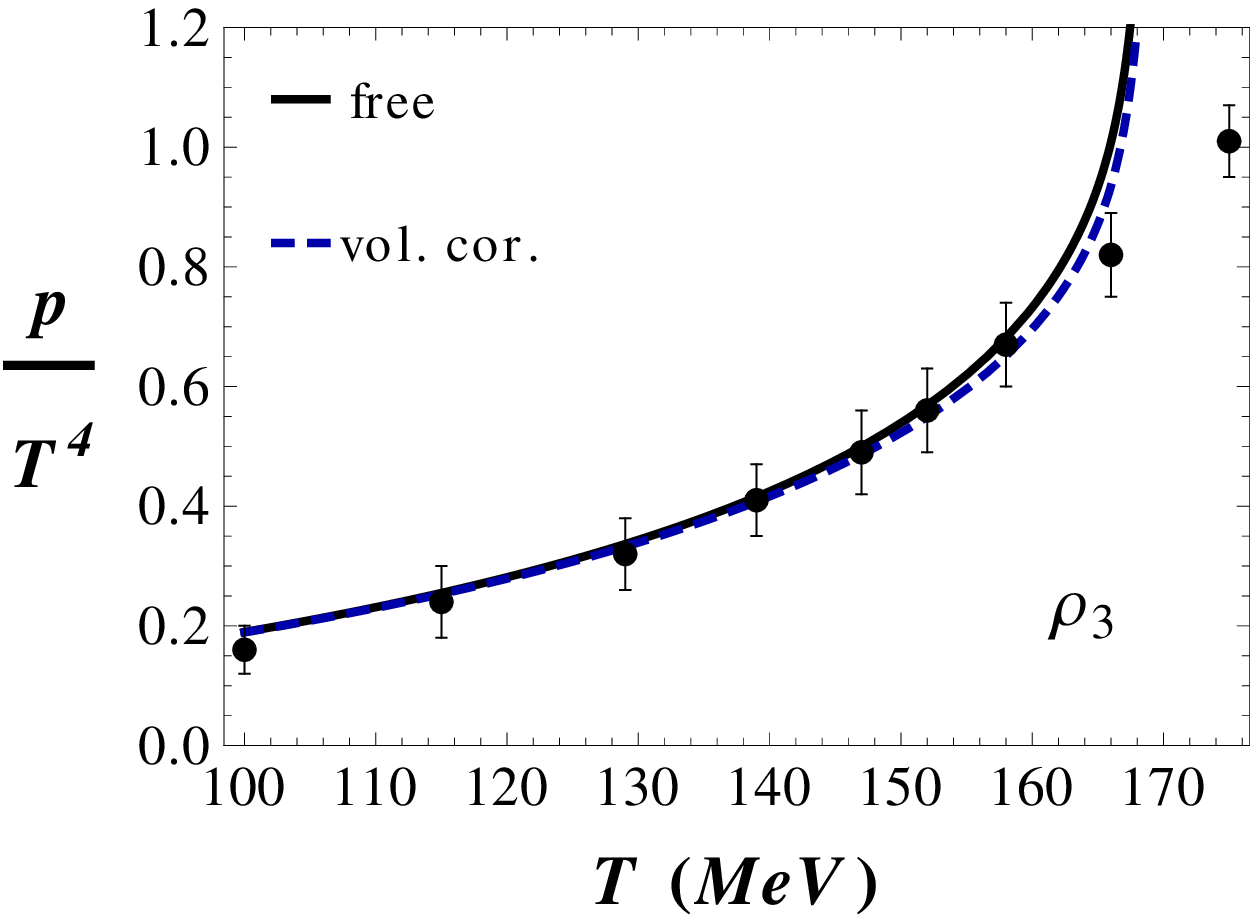,width=0.6\linewidth,clip=}\\
\epsfig{file=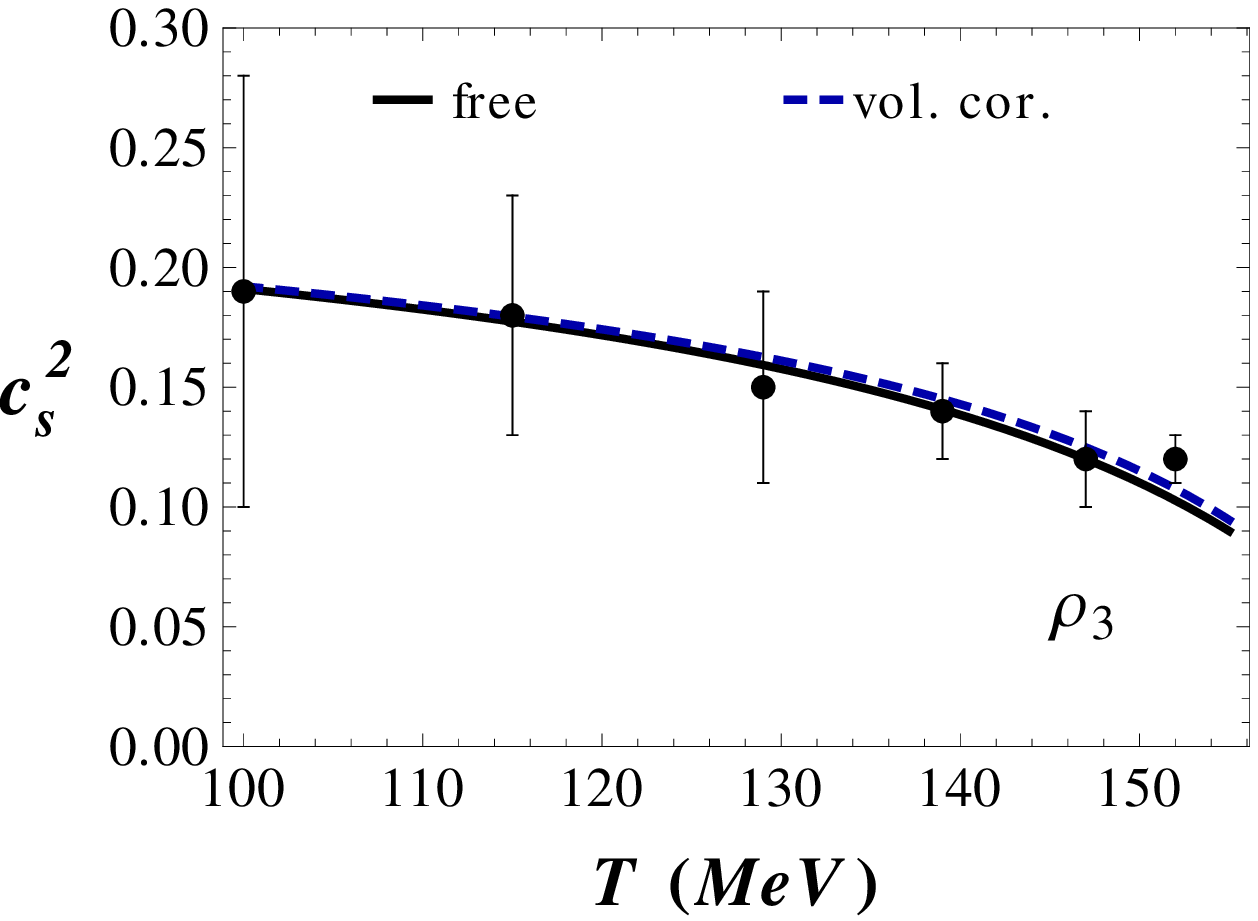,width=0.6\linewidth,clip=}
\caption{Trace anomaly, pressure, and speed of sound squared for the hadron resonance model with density of states $\rho_3$ and an excluded volume radius of $r=0.2$ fm. The dashed blue curves denote the excluded volume results 
while the solid black curves show the corresponding quantities without excluded volume corrections. The data points correspond to the $N_t=10$ lattice data published in
Ref.\ \cite{Borsanyi:2010cj} (obtained from table 5 in that paper).} \label{thermoplotrho3vdw}
\end{figure}

Given that the free hadronic gas of the previous section provided a good description of the data, one should expect that the volume corrections should be minimal in this case. In fact, one can again use the lattice data and 
the different hadron mass spectra discussed before to show that the excluded radius cannot be larger than $0.2$ fm for $\rho_i$ ($i=1,2,3$). $\rho_4$ is not considered in this case since it fails to describe the data 
for $T>140$ MeV. We show a comparison between the lattice data and the $\rho_3$ model curves for $r=0.2$ fm in Fig.\ \ref{thermoplotrho3vdw}. The small excluded volume shifts the curves slightly downwards, which makes them get
closer to the lattice data at $T=160$ MeV. Similar results hold for the other exponentially increasing spectra considered before. For larger excluded volumes the hadron resonance gas curves start to deviate from 
the lattice at lower temperatures and this is why we here take $r\leq 0.2$ fm. This analysis shows that volume corrections do not play a significant role in the description of lattice data (at least for the mass spectra 
parameters determined in the previous section). Of course, given the known uncertainties in lattice calculations at low temperatures, another possibility would be to define the parameters in a way that the model including the continuous spectrum 
fits the lattice data only at higher temperatures around $T\sim 150$ MeV, as it was done in \cite{NoronhaHostler:2008ju}. This will be discussed further in the next section.    
    
\section{Calculation of the Shear Viscosity to Entropy Density Ratio and The Shear Relaxation Time Coefficient of the Hadron Resonance Gas Model with Excluded Volume Corrections}

The computation of the transport properties of a hadronic mixture is not an easy task. There have been several studies on this subject in the last few years \cite{Prakash:1993kd,Gorenstein:2007mw,Itakura:2007mx,NoronhaHostler:2008ju,bass,pal,Chakraborty:2010fr}.
In order to find at least an estimate of the order of magnitude of the $\eta/s$ ratio of hot hadronic matter at $T\sim 160$ MeV, we follow the approximations made by Ref.\ \cite{Gorenstein:2007mw} where the shear viscosity of a multi-component gas of hadrons
and resonances in the excluded volume approximation (described in the previous section) is given by
\begin{equation}
 \eta= \frac{5}{64 \,r^2}\left(\frac{T}{\pi}\right)^{1/2} \frac{T}{2\pi^2\,n(T)}\int_0^\infty dm\,\rho(m)\,m^{5/2} K_{5/2}\left(\frac{m}{T}\right)\,.
\label{sheargorenstein} 
\end{equation}
Currently, it is not known how to compute the contribution to the shear viscosity from heavily massive and highly unstable resonances that cannot rigorously be described using the Boltzmann equation. These states contribute significantly to the thermodynamic properties
of the matter at high tempetatures (as shown in the previous sections) and, due to their rapid decay, it is natural to assume that their presence will affect the mean free paths of the other hadrons. In \cite{NoronhaHostler:2008ju}, it was assumed that the 
mean free path of these resonances with $m > 2$ GeV equals their inverse decay width. Obviously, further studies have to be carried out to properly include the effects of Hagedorn states on transport coefficients of hot hadronic matter.  
While the formula in Eq.\ (\ref{sheargorenstein}) may only provide an estimate of the shear viscosity of an interacting hadron gas, the temperature behavior of $\eta$ computed with this approximation \cite{Gorenstein:2007mw} 
follows the estimates made using other methods and thus we shall proceed using this formula.

\begin{figure}
\centering
\epsfig{file=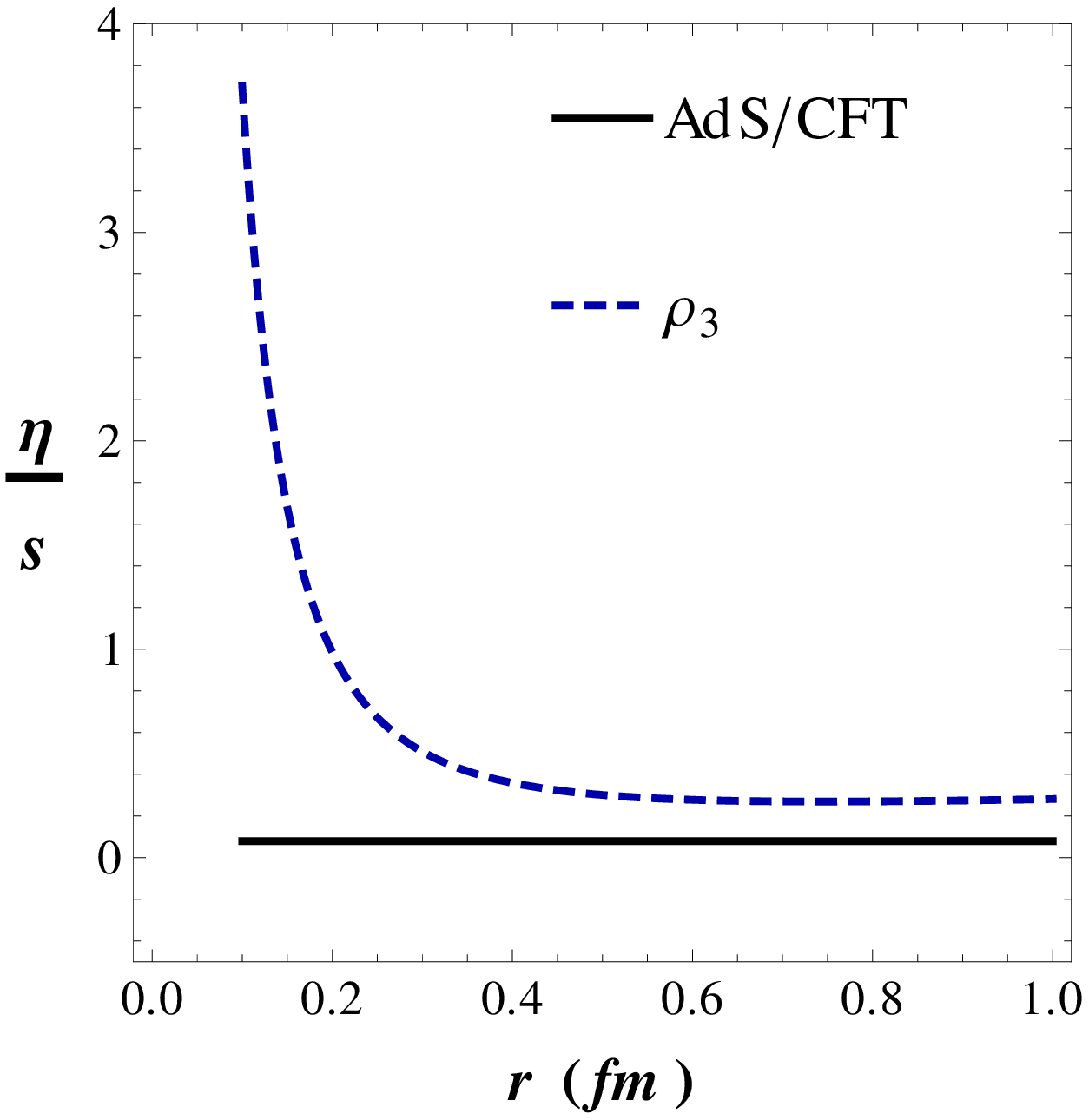,width=0.6\linewidth,clip=}
\caption{The ratio $\eta/s$ (dashed blue line) as a function of the excluded volume radius computed via (\ref{sheargorenstein}) using the density of states $\rho_3$ in Eq.\ (\ref{rhofrautschi}) and $T=155$ MeV. The black line denotes
the viscosity lower bound $\eta/s = 1/(4\pi)$.} \label{plotetasrho3varyr}
\end{figure}

In Eq.\ (\ref{sheargorenstein}), the dependence on the excluded radius only appears via the $1/r^2$ factor. However, when computing $\eta/s$, the entropy density should be the one determined within the same approximation, i.e, $s_v(T)$. Therefore, 
even in this approximation $\eta/s$ possesses a nontrivial dependence on the hadron cross section $\sim r^2$. This is shown in Fig.\ \ref{plotetasrho3varyr} where the $\eta/s$ ratio is computed for different $r$'s using $\rho_3$ (\ref{rhofrautschi}) and $T=155$ MeV. 
Note that for the highest temperature considered $T=155$ MeV, $\eta/s$ depends very weakly on $r$ when the excluded radius is larger than 0.2 fm (using this specific set of parameters that define the mass spectrum).

\begin{figure}
\centering
\epsfig{file=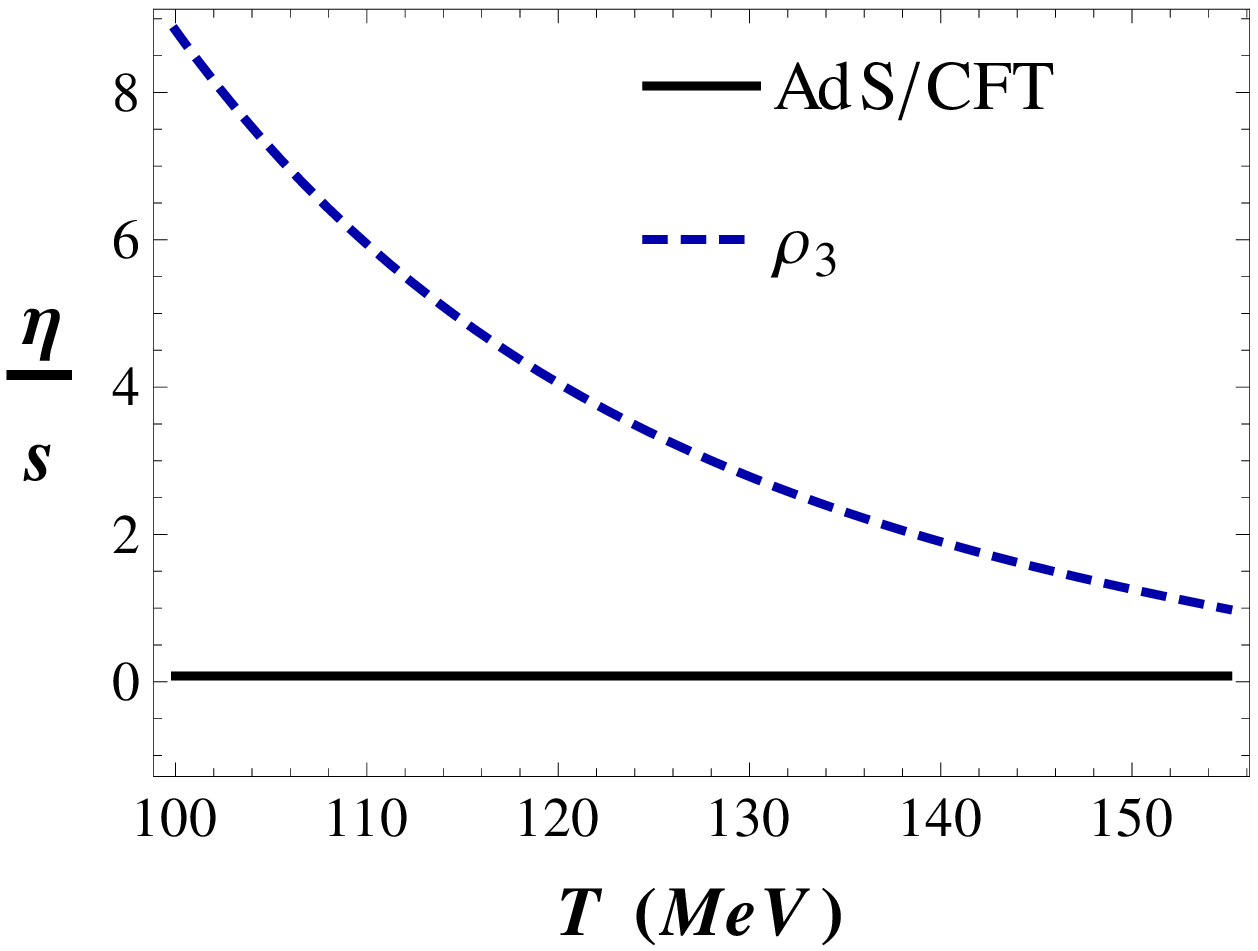,width=0.6\linewidth,clip=}
\caption{The ratio $\eta/s$ (dashed blue line) computed within the excluded volume approximation defined via Eqs.\ (\ref{vdwpressure}) and (\ref{sheargorenstein}) using the density of states $\rho_3$ in Eq.\ (\ref{rhofrautschi}) and $r=0.2$ fm. 
The entropy density computed in this model fits the lattice data \cite{Borsanyi:2010cj} in the temperature range shown. The black line denotes the viscosity lower bound $\eta/s = 1/(4\pi)$.} \label{plotetasrho3}
\end{figure}

In Fig.\ \ref{plotetasrho3} we show the temperature dependence of $\eta/s$ computed using the density of 
states $\rho_3$ in Eq.\ (\ref{rhofrautschi}) and $r=0.2$ fm. As discussed in the previous section, for the specific choice of parameters that define $\rho_3$, calculations of thermodynamical quantities performed with an excluded radius of $r=0.2$ fm can describe
the lattice data in the entire temperature range, $T=100-160$ MeV. This value of the excluded radius is smaller than other estimates \cite{Gorenstein:2007mw}, which however were not constrained by fitting the lattice data. Perhaps it would be more physical
to consider a model where the excluded radius increases with the mass of the hadron but for simplicity's sake in the current we will limit ourselves to a constant excluded radius hoping that we are correct within an order of magnitude.  
In Fig.\ \ref{plotetasrho3}, we see that the $\eta/s$ ratio remains an order of magnitude above the viscosity lower bound up to $T=155$ MeV (this remains the case when other expressions for the density of states mentioned
in the previous sections are used). The entropy density computed in this case matches the lattice data well, as it can be inferred from the other quantities shown in Fig.\ (\ref{thermoplotrho3vdw}). 
Therefore, this simple hadron resonance gas model with constant excluded volume corrections is able to describe the thermodynamic quantities computed by lattice data below $T=160$ MeV and the corresponding $\eta$ is computed self-consistently 
within the same framework. 

\begin{figure}
\centering
\epsfig{file=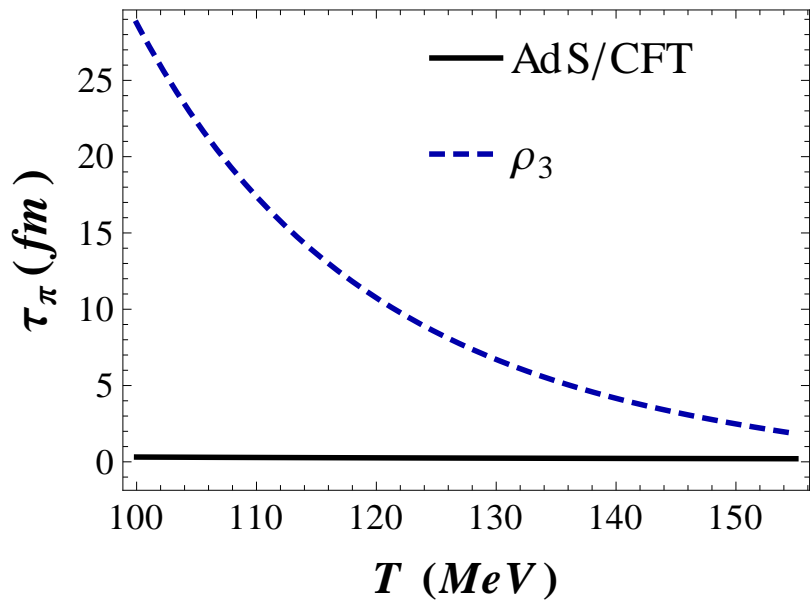,width=0.6\linewidth,clip=}
\caption{The lowest relaxation time coefficient $\tau_\pi$ computed using the inequality in Eq.\ (\ref{tomoi}) for the hadron resonance gas model defined with density of states $\rho_3$ and $\eta/s$ and $c_s^2$ computed within the 
excluded volume approximation with excluded volume radius $r=0.2$ fm (dashed blue curve). The black line denotes the lowest value for $\tau_\pi$ in a conformal ``nearly perfect'' fluid where $c_s^2=1/3$ and $\eta/s=1/(4\pi)$.} \label{plottaupirho3}
\end{figure}

In viscous hydrodynamic calculations of the QGP time evolution \cite{bjorn,song,romatschke}, there is at least another transport coefficient that must be included in the fluid equations, the shear viscosity relaxation time $\tau_\pi$, which 
enters in second order viscous hydrodynamic calculations. In fact, in relativistic fluids causality is intimately connected to stability \cite%
{hiscock,us} and Israel and Stewart \cite{IS} were among the first to
understand that the characteristic times within which fluid dynamical
dissipative currents relax towards their asymptotic Navier-Stokes values cannot be arbitrarily small. Using the Boltzmann equation, it is possible to show \cite{IS,dkr,Denicol:2012cn,Denicol:2011fa} that in
relativistic gases $\tau _{\pi}$ is of the order of the microscopic collision time. A detailed linear stability analysis made in Ref.\ \cite{Pu:2009fj} showed that stability and causality require that $\tau_\pi$ in any viscous relativistic fluid cannot be 
arbitrarily small. In fact, this transport coefficient must obey the following inequality \cite{Pu:2009fj}
\begin{equation}
 \tau_\pi \geq \frac{4}{3}\frac{\eta}{s \,T}\frac{1}{(1-c_s^2)} \,.
\label{tomoi}
\end{equation}
While we shall not compute $\tau_\pi$ for the hadron resonance gas model considered here, we find it instructive to consider the smallest $\tau_\pi$ implied by the inequality above since it provides an estimate for the value of this parameter that can
be used in hydrodynamic simulations. In Fig.\ (\ref{plottaupirho3}) we show (dashed blue line) the lowest value for $\tau_\pi $ computed using the $\eta/s$ in Fig.\ \ref{plotetasrho3} that fulfills the stability and causality criteria. 
We also show in the same plot the lowest value for $\tau_\pi$ in a generic conformal ``nearly perfect'' fluid where $c_s^2=1/3$ and $\eta/s=1/(4\pi)$ for reference. Note that the lowest value of $\tau_\pi$ computed in the hadron resonance gas 
model employed here remains well above the lowest value for a nearly perfect
fluid up to $T=160$ MeV.    

\begin{figure}
\centering
\epsfig{file=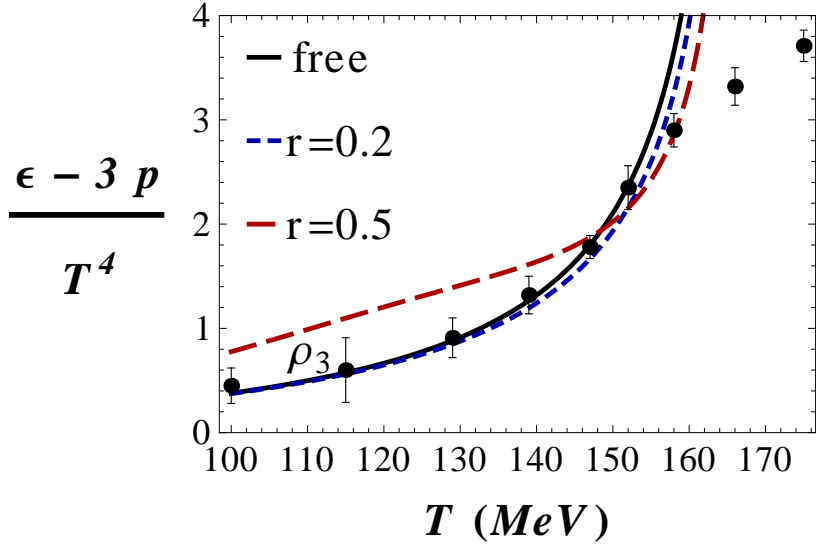,width=0.6\linewidth,clip=}
\caption{Comparison between the model's trace anomaly computed different excluded volumes and the $N_t=10$ lattice data published in
Ref.\ \cite{Borsanyi:2010cj} (obtained from table 5 in that paper). The solid black curve was computed using $\rho_3$ with the parameters defined in the table \ref{tab:par}. The short-dashed blue curve is computed using the same $\rho_3$ but with an excluded volume
of $r=0.2$ fm. The long-dashed red curve was computed using $\rho_3$ but with $A_3=3.7$ GeV$^2$ and $m_{03}=1$ GeV together with an excluded volume of $r=0.5$ fm.} \label{plote3pnewrho3}
\end{figure}

\begin{figure}
\centering
\epsfig{file=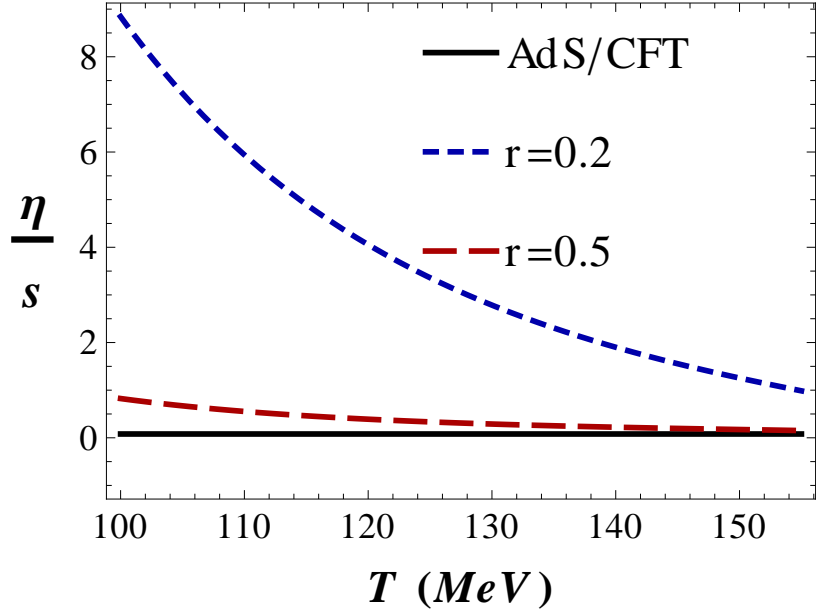,width=0.6\linewidth,clip=}
\caption{The ratio $\eta/s$ computed within the excluded volume approximation. The short-dashed blue curve is computed using $\rho_3$ with the parameters defined in the table \ref{tab:par} but with an excluded volume
of $r=0.2$ fm. The long-dashed red curve was computed using $\rho_3$ but with $A_3=3.7$ GeV$^2$ and $m_{03}=1$ GeV together with an excluded volume of $r=0.5$ fm.} \label{etasnew}
\end{figure}

\begin{figure}
\centering
\epsfig{file=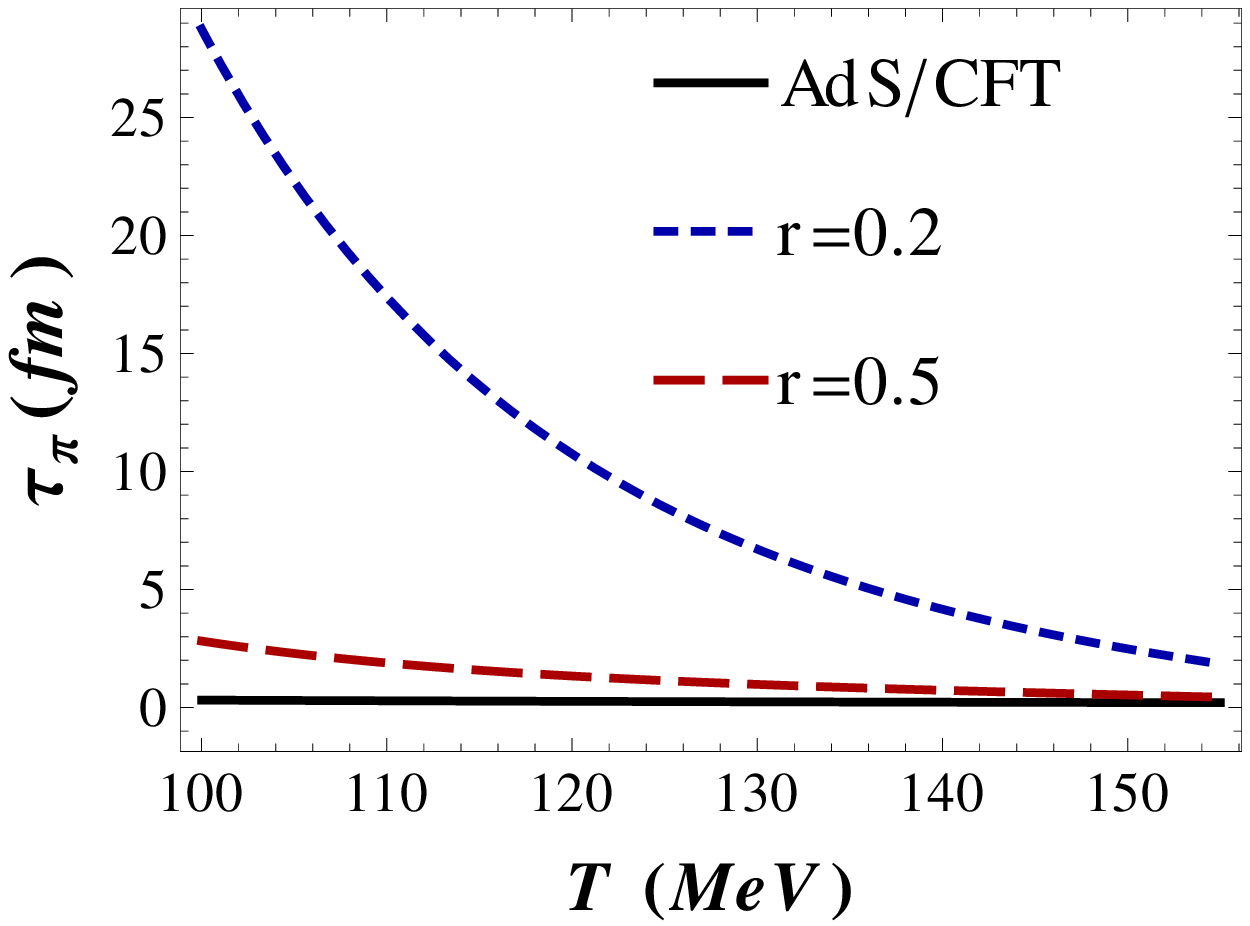,width=0.6\linewidth,clip=}
\caption{The lowest relaxation time coefficient $\tau_\pi$ computed using the inequality in Eq.\ (\ref{tomoi}) for the hadron resonance gas model. The short-dashed blue curve is computed using $\rho_3$ with the parameters defined in 
the table \ref{tab:par} but with an excluded volume of $r=0.2$ fm. The long-dashed red curve was computed using $\rho_3$ but with $A_3=3.7$ GeV$^2$ and $m_{03}=1$ GeV together with an excluded volume of $r=0.5$ fm.}
 \label{taupinew}
\end{figure}

Throughout our calculations thus far we have the underlying assumption that the hadron gas model must fit the entire lower temperature region $T\sim 100$ MeV of the lattice data. However, it is interesting to consider the possibility of just fitting
the lattice data at higher temperatures $T\sim 150$ MeV to see how that affects the $\eta/s$ calculation. This can be accomplished by changing some of the parameters of the hadron mass spectrum. Before doing so, we increase $r=0.5$ fm \cite{Gorenstein:2007mw}. Then
we refitted the parameters $A_3$ and $m_{03}$ in (\ref{rhofrautschi}) in order to fit only the high temperature region of the lattice trace anomaly. This is shown in Fig.\ (\ref{plote3pnewrho3}). The corresponding $\eta/s$ and shear relaxation
time computed with this setup are shown in Figs.\ (\ref{etasnew}) and (\ref{taupinew}). Note that with this new set of parameters $\eta/s$ drops down to $1/(4\pi)$ around $T=160$ MeV and $\tau_{\pi}$ also decreases significantly in that region. The decrease of
the transport coefficients is mainly driven by the larger entropy that results from this new fit that is only constrained by the higher lattice temperatures. This emphasizes the sensitivity of such calculations to the thermodynamic properties of the matter.

\section{Conclusions}

In this paper we performed a detailed comparison of hadron resonance gas model calculations to recent lattice data \cite{Borsanyi:2010cj} that confirmed the need for an exponentially increasing density of hadronic states. Also, we showed that the hadron mass spectrum, extracted from a comparison to lattice data, is compatible with $\rho(m) \sim m^{-a}\exp(m/T_H)$ at large $m$ with $a> 5/2$ (where $T_H \sim 167$ MeV). With this specific $m^{-a}$ (with $a>5/2$) factor in the density of states, heavy resonances most likely undergo 2-body decay (instead of multi-particle decay) \cite{Frautschi:1971ij}, which facilitates their inclusion into hadron transport codes. 
Moreover, we have computed the shear viscosity to entropy density ratio of this system within the excluded volume approximation and the results suggest that $\eta/s$ of hot 
hadronic matter is very sensitive to the temperature dependence of the thermodynamic quantities. Using this calculation of $\eta/s$, we were able to compute the lowest value for the shear 
relaxation time coefficient used in second order hydrodynamic calculations that respects the criteria of causality and stability of a relativistic viscous fluid. 

Previous estimates for the $\eta/s$ ratio in a hadronic gas \cite{NoronhaHostler:2008ju} had concluded that hadronic matter at temperatures $T \sim 190$ MeV behaved as a nearly perfect fluid. This is not at odds with the findings presented in this
paper as we explained in the previous section. In fact, the curve shown in Fig.\ (\ref{plotetasrho3}) approaches $1/(4\pi)$ when continued to temperatures $\sim 190$ MeV. Moreover, if only the high temperature region of the lattice data is fitted, then the transport
coefficients calculated here decrease significantly (e.g., $\eta/s \sim 1/(4\pi)$ near $T = 160$ MeV). This highlights the importance of knowning the correct temperature dependence of the thermodynamic quantities of QCD at temperatures $\sim 100-160$ MeV since it
may play a important role in the calculation of transport coefficients.

The key difference between these studies is the lattice data used as a reference for the hadron gas calculations. 
Ref.\ \cite{NoronhaHostler:2008ju} used the most recent lattice data at the time \cite{Cheng:2007jq} which indicated a phase transition pseudo-critical temperature $T_c\sim 196$ MeV. The much lower value for this pseudo-critical temperature found 
in \cite{Borsanyi:2010cj} severely reduced the value of the maximum temperature at which the hadron resonance gas is still applicable since ``$T_c$'' decreased from 190 MeV to 160 MeV. Given that this low pseudo-critical temperature has already
been independently confirmed by other lattice groups \cite{Bazavov:2011nk}, if there is no change in the low temperature behavior of the thermodynamic quantities as determined by lattice, the analysis performed in this paper indicates 
that the hot hadronic matter formed in ultrarelativistic heavy ion collisions is far from being a nearly perfect fluid. However, this should be taken with a grain of salt given the sensitivity mentioned above within the transport coefficients. 

Since $(\varepsilon-3p)/T^4$ in the lattice data \cite{Borsanyi:2010cj} continues to increase 
until it reaches a turning point around $T\sim 200$ MeV, one may wonder if it is possible to devise a model that reduces to the Hagedorn resonance gas discussed here at low temperatures 
while also incorporates the correct degrees of freedom in the crossover region between $T\sim 160-200$ MeV. Perhaps such an effective model can be constructed by taking into account the Polyakov loop \cite{Dumitru:2010mj}.  

This work was partially supported by the Helmholtz International Center for FAIR
within the framework of the LOEWE program launched by the State of Hesse. J.~Noronha-Hostler is supported by Fundacao de Amparo a Pesquisa do Estado de Sao Paulo (FAPESP). J.~Noronha thanks Conselho Nacional de 
Desenvolvimento Cientifico e Tecnologico (CNPq) and Fundacao de Amparo a Pesquisa do Estado de Sao Paulo (FAPESP) for financial support.

\end{document}